# A native signed incoherent photonic matrix core on thin-film lithium niobate supporting in situ backpropagation

Yuan Ren[1,2], Yong Zheng[2]*, Ruixue Liu[1,2], Min Wang[2] and Ya Cheng[1,2,3,4,5,]*

[1]State Key Laboratory of Precision Spectroscopy, East China Normal University, Shanghai 200062, China.

[2]The Extreme Optoelectromechanics Laboratory (XXL), School of Physics, East China Normal University, Shanghai 200241, China.

[3]Hefei National Laboratory, Hefei 230088, China.

[4]Shanghai Research Center for Quantum Sciences, Shanghai 201315, China.

[5]Collaborative Innovation Center of Extreme Optics, Shanxi University, Taiyuan 030006, China.

[*]Corresponding authors. Email: Yong Zheng (yzheng@phy.ecnu.edu.cn),

Ya Cheng (ya.cheng@siom.ac.cn).

**Abstract**

Artificial intelligence workloads increasingly demand computing architectures combining high throughput, energy efficiency, and physical scalability. Here we present a signed incoherent optical matrix multiplier on thin-film lithium niobate that natively supports signed inputs and weights. We demonstrate closed-loop in situ backpropagation on a 16 × 16 photonic core, including forward computation, nonlinear operations, error propagation, and gradient computation. Measured physical outputs directly participate in optimization, thereby incorporating actual device responses and nonidealities into training. The core achieves 8-bit multiplication and 10-bit accumulation precision across 256 channels, maintains 10-bit accuracy in tiled 256 × 256 matrix computation, and further executes Transformer linear operations in a BERT-mini workload. This work presents a fundamental, scalable building block that addresses key limitations in practical optical computing, thereby opening avenues for large-scale, high-efficiency photonic neuromorphic systems.

## Introduction

Artificial intelligence (AI) models repeatedly execute signed linear transformations during both inference and parameter updating in training (*1-3*). In digital accelerators, the associated matrix multiply-accumulate (MAC) operations consume energy both in arithmetic computation and in the repeated movement of activations and weights between memory, compute units and interfaces. Amid the escalating computational demands of AI, physical neural networks that leverage physical laws to inherently perform neural computations and potentially alleviate the energy and speed limitations of electronics have emerged as a promising beyond von-Neumann paradigm (*4,5*). Among diverse realizations, optical neural networks (ONNs) distinguish themselves by capitalizing on the innate properties of light, including high speed, low latency, low power consumption, and high parallelism. These characteristics offer the potential for high energy efficiency and throughput in large-scale parallel computation (*6-11*).

Among different ONN architectures, incoherent intensity-domain MAC computing provides a particularly favorable route toward scaling because it encodes computation through local optical power transfer functions (*12-14*) rather than relying on phase-coherent interference across the Mach-Zehnder interferometer (MZI) meshes (*15-21*), thereby relaxing the requirements for global phase stabilization and calibration. However, due to the inherently non-negative nature of optical intensity, most existing incoherent architectures cannot simultaneously process positive and negative inputs and weights without additional modulation channels, duplicated optical paths or repeated detection (*14,22-23*). This limitation is particularly restrictive for neural network training. During training, intermediate activations can be signed, whereas the errors and gradients propagated during backpropagation inherently carry both positive and negative values (*2,24-27*). If negative quantities cannot be represented directly in the photonic domain, part of the training procedure must be reformulated, implemented using duplicated hardware, or transferred to electronic computation (*28-30*). Therefore, native four-quadrant multiplication does not merely expand the numerical range but constitutes a key capability for a photonic processor that physically executes backpropagation. A second challenge is numerical stability under repeated computation. Optical neural networks are analog systems, and high precision in a single matrix multiplication does not guarantee comparable fidelity after repeated operations, tiled matrix execution, or multilayer propagation. Small errors arising from modulation, detection, and accumulation can propagate across successive computational stages. Whether a finite photonic core can preserve its precision over many computational cycles and when implementing matrices much larger than its native dimension through tiling is therefore critical to practical scalability. Together, native signed computation and precision retention under repeated reuse constitute two central requirements for scaling incoherent photonic processors from individual operations to physical training and large-matrix workloads.

Here we demonstrate a fully programmable 16 × 16 thin-film lithium niobate (TFLN) signed incoherent photonic matrix core that addresses these challenges within a unified architecture. Without duplicating the programmable weight array or requiring repeated detection, the processor performs four-quadrant multiplication between signed inputs and signed weights using calibrated wavelength pairs that traverse the same programmable array, followed by differential readout. This capability enables forward propagation, PReLU activation and its derivative, the derivative of the mean squared error (MSE) loss function, error backpropagation and gradient computation to be implemented directly on the photonic chip. We further deploy this chip to train a noisy extended

XOR classifier, thereby experimentally demonstrating closed-loop on-chip in situ backpropagation in an incoherent MZI array. Because the training loop directly uses outputs measured from the physical core, the realized device responses and associated nonidealities are incorporated into the optimization process. All 256 computing channels are experimentally characterized, with precision exceeding 8 bits across 1.28 million signed multiplications and accumulated precision exceeding 10 bits. By repeatedly reusing the same 16 × 16 core, we perform 256 × 256 matrix computation while retaining approximately 10-bit precision, and maintain this precision over 60 cycles of 64-dimensional matrix computation. The same physical core is further reused via tiled execution for a BERT-mini workload on the AG News dataset, with inference accuracy on par with conventional digital baselines. The front-end lithium niobate modulators further provide a measured 50 GHz electro-optic bandwidth for high-rate input streaming. Together, this study presents a scalable signed building block for optical matrix multiplication, opening the door to future expansions in large-scale practical ONNs. It is noteworthy that before we have demonstrated a photonic neural network on TFLN which is named ZEN-1 (ZEN: zero energy-consumption neural-network) (21), thus the current photonic matrix core is named ZEN-2 which can be considered as a progressive milestone in the up-scaling of ZEN until one day such high-performance photonic neural networks can be used in practical circumstance.

## Results

### A signed incoherent matrix core for in situ backpropagation

The processor comprises a 16 × 16 signed optical matrix core (Fig. 1A) which was fabricated on TFLN using photolithography-assisted chemo-mechanical etching (PLACE), a waveguide process that offers low-roughness sidewalls for ultra-low-loss waveguides (*31-33*) and supports large-scale repetitive interferometric cells (*34-35*) (Fig. S1). The core integrates 32 high-speed input modulators encoding the signed input vector **X**, 16 asymmetric MZIs for routing the calibrated wavelength pairs, and 256 programmable weight elements implementing the signed weight matrix **W** updated at a lower cadence, enabling optical matrix multiplication **Y = WX**. Building on our previous work (*36*), the core elevates dual-wavelength differential multiplication from a scalar principle to a reusable spatial matrix primitive. The scheme realizes four-quadrant operation through wavelength assisted differential encoding (Fig. 1B). For a bipolar input vector $\boldsymbol{X} \in [-1,1]^N$, each signed element $X_i$ is represented by a pair of non-negative components $x_{i+}$ and $x_{i-}$, satisfying $X_i = x_{i+} - x_{i-}$. The two components are independently mapped onto optical carriers at $\lambda_1$ and $\lambda_2$, via two high-speed electro-optic modulators biased at their working points. The modulated carriers subsequently enter an asymmetric MZI wavelength routing stage. Its arm-length difference $\Delta L$ is carefully designed so that the $\lambda_1$ and $\lambda_2$ signals launched from different input ports are directed toward a common output port. This combined signal is subsequently split and fed into an array of weight multiplication units, each implemented with an unbalanced MZI identical to that in the multiplexing module. As a result, the applied weights $w_{ij+}$ and $w_{ij-}$ for $\lambda_1$ and $\lambda_2$ are complementary, satisfying $w_{ij+} + w_{ij-} = 1$. The two complementary output ports of the weighting interferometer are detected. After normalization, differential detection of the two output ports yields $(w_{ij+} - w_{ij-})(x_{i+} - x_{i-}) = W_{ij}X_i$, where $W_{ij} = w_{ij+} - w_{ij-} \in [-1,1]$ (Fig. S2).

At the layer level, the matrix core implements the three signed linear relations used by backpropagation. The forward transformation for layer $l$ is

$$\mathbf{Y}^{(l)} = \mathbf{W}^{(l)}\mathbf{X}^{(l)} + \mathbf{b}^{(l)} \tag{1}$$

After a local activation, the error is propagated through the transposed matrix,

$$\mathbf{g}^{(l-1)} = \left[\mathbf{W}^{(l)}\right]^{\mathsf{T}}\boldsymbol{\delta}^{(l)} \tag{2}$$

$$\boldsymbol{\delta}^{(l-1)} = \mathbf{g}^{(l-1)} \odot f'^{(l-1)}\left(\mathbf{Y}^{(l-1)}\right) \tag{3}$$

where $\boldsymbol{\delta}^{(l)}$ is the error of the $l$th layer and $\odot$ denotes elementwise multiplication. The corresponding weight-gradient contribution is

$$\frac{\partial \mathcal{L}}{\partial \boldsymbol{W}^{(l)}} = \boldsymbol{\delta}^{(l)}\left[\boldsymbol{X}^{(l)}\right]^{T} \tag{4}$$

These relations are placed within the layer-wise learning flow (Fig. 1C), and their linear portions are assigned to the physical chip (Fig. 1D). The programmed array performs the forward multiplication and the transposed signed linear operation for error propagation, while the same dual-wavelength multiplication mechanism measures the signed products required for gradient formation. Task-specific optical, electronic or hybrid modules supply nonlinear functions, loss construction and update rules. The controller assembles the measured gradient-related products to obtain the batch-averaged gradients before updating the programmed weights. This single physical matrix core is thereby reused across the signed forward, backward and gradient operations that define the learning dataflow (Method S1.2).

**Electro-optic characterization of input and weight modulators**

The physical mapping (Fig. 1) naturally separates the processor into a high-speed input path and a comparatively slow programmable weight path, reflecting the data-flow characteristics of neural network computation (*37-38*). During inference, a trained weight matrix remains fixed while many input samples or activation vectors are successively processed. During training, the same weights are likewise reused across the samples within a mini-batch and are updated only after the corresponding gradients have been accumulated. The input variables must therefore be refreshed much more frequently than the matrix weights. Under such a weight-stationary computing scheme, a higher input-modulation rate directly increases the number of activation vectors that can be processed by the same programmed matrix per unit time, making high-speed electro-optic modulation particularly advantageous for large batch and high throughput neural computation.

We therefore designed the input region for high-speed variable loading and characterized its electro-optic response. At 1550 nm, a representative input modulator exhibited a fitted half-wave voltage of 3.36 V (Fig. 2A). Two representative input branches showed 3-dB electro-optic bandwidths exceeding 50 GHz (Fig. 2D), enabling rapid streaming of input variables into the photonic matrix core. At a symbol rate of 50 Gbaud, the dual-wavelength 16 × 16 matrix core corresponds to a peak processing rate of 51.2 TOPS (Supplementary text S2.6.4).

The weight region, by contrast, is optimized for stable matrix programming rather than high-speed data streaming. During inference, a programmed weight state can be retained while a large number of input vectors are processed; during training, each programmed state can likewise be reused across multiple samples before reconfiguration for subsequent operations or parameter updates. We therefore characterized the wavelength-dependent weight response required for differential signed encoding. At 1546.7 and 1550 nm, the two optical channels exhibited complementary transfer characteristics over the programmed range (Fig. 2B). A single electrical weight setting consequently produces opposite transmission changes at the two wavelengths, enabling differential representation of signed weights. No measurable electro-optic drift was observed over two hours, during which the differential output error showed a standard deviation of only 0.05% (Fig. 2C). This stability supports retention of the programmed weight state between updates, as required for the low-update-rate weight path.

**Precision of signed multiplication and matrix operations**

To evaluate the precision and scalability of the signed photonic computing core, we first performed comprehensive calibration of the 256 computational channels. The electrical control and acquisition sequence, device-specific transfer functions, detector normalization and differential output offsets were individually calibrated before array-level measurements (Method S1.3 and S1.4). All sixteen 1 × 16 computational groups were independently characterized to verify operation across the entire array.

Following calibration, the complete array was evaluated using signed input–weight combinations covering the full bipolar operating range. Across 1280 independent measurement sets, each containing 1000 input–weight combinations, the signed multiplication errors were symmetrically distributed around zero (Fig. 3A).

Measuring the combined precision contributions from electrical signal generation, acquisition and photonic computation, the root-mean-square error (RMSE) corresponded to an effective number of bits (ENOB) of 8.219 bits over the full signed output range. The best-performing 1000-point measurement set achieved 9.244-bit precision. Further statistical analysis across the 256 physical computing channels yielded an average precision of 8.291 bits with a standard deviation of 0.144 bits (Fig. 3B). The pointwise distribution is visualized (Fig. 3C), with the error metric, ENOB definition, and normalization provided in Supplementary Text S2.2. The same calibration procedure and four-quadrant computing mechanism were subsequently applied in the on-chip training experiments to process signed activations, errors and gradient-related quantities in the noisy XOR task.

We next examined the precision of matrix operations when the physical 16 × 16 core was reused to perform larger-scale computations. Using matrix tiling, the calibrated core was applied to matrix–vector and matrix–matrix operations with increasing dimensions. The output precision remained close to the native core performance during scaling, reaching 9.87-bit MAC precision for tiled 256 × 256 matrix computation (Fig. 3D). The higher normalized MAC precision is consistent with partial cancellation of positive and negative product errors during accumulation, with the summed error normalized by the sum of the absolute weights in the corresponding output row (Supplementary text S2.2). This tiled operation provides the computational basis for mapping neural-network layers whose matrix dimensions exceed the physical size of the photonic core.

Finally, we evaluated long-sequence computation precision by repeatedly applying the same signed matrix operation over 60 consecutive computation cycles. The per-output MAC precision remained stable throughout the entire sequence, with an average precision of 9.84 bits and a final-cycle precision of 9.64 bits; the corresponding individual multiplication precision remained 8.58 bits (Fig. 3E). This repeated computation stability enables the same photonic core to be reused across successive computational stages, which is required for the multi-layer matrix operations involved in the BERT-mini workload.

**Closed-loop XOR training with the photonic core**

With stable precision, the four-quadrant photonic computing architecture enables signed information to propagate through both forward and backward computational paths. We therefore implemented a closed-loop in situ backpropagation experiment for a noisy nonlinear XOR classifier (Fig. 4A). The two-layer neural network was mapped onto the 16 × 16 photonic core, where matrix operations involved in the forward pass, error propagation and gradient-related product measurement were performed using the same physical signed computing mechanism. The measured optical outputs and error signals were directly used within the training loop to determine parameter updates.

The nonlinear operations required for training were also incorporated into the photonic processor. The PReLU activation function was implemented by separating positive and negative branches and applying calibrated optical transfer responses (Fig. 4B). For the mean-squared-error objective, the output-layer error signal was calculated from the measured network output and target vector,

$$\boldsymbol{\delta}^{(2)} = \frac{2}{N}(\mathbf{Z} - \mathbf{Y}_{\text{label}}) \odot f'_{\text{out}}\left(\mathbf{Y}^{(2)}\right) \tag{5}$$

where $N$ is the batch size, $\mathbf{Z}$ is the network output and $\mathbf{Y}_{\text{label}}$ is the target vector (Fig. 4C). The complete dual-wavelength experimental sequence, gradient definitions and element-resolved trajectories are given in Supplementary Text S2.3.

Using the on-chip forward propagation, nonlinear activation, error evaluation and backpropagation processes, the photonic processor formed a closed-loop training system and trained the XOR dataset with 200 samples over 80 epochs. The classification accuracy reached 0.940, while the mean-squared-error loss decreased to 0.0588 (Fig. 4E). The resulting decision boundary correctly separated the two nonlinear XOR classes (Fig. 4D). During training, the element-resolved gradients exhibited both positive and negative values (Fig. S6), while the root-mean-square magnitudes of the gradient groups decreased toward smaller residual values (Fig. 4F), consistent with the signed updates used in network optimization.

**Scaling a photonic core to Transformer linear operations**

The XOR experiment established a closed-loop training workflow based on signed photonic computation. We next investigated whether the 16 × 16 photonic core could be extended through tiled execution to perform the large-scale linear operations underlying Transformer models. We mapped the linear operations of a four-layer BERT-mini encoder trained on the AG News dataset onto the photonic processor through tiled execution (Fig. 5A, Fig. S7). The photonic processor executed the major matrix operations in the Transformer workload, including the query, key and value projections, attention score calculation, multiplication of the attention probabilities by the value matrix, attention output projection, feed-forward layers and classification head. These operations are decomposed into 16 × 16 submatrices and sequentially scheduled on the photonic core. Tokenization, embedding lookup, bias addition, softmax, GELU activation, residual connections and layer normalization were performed electronically. For the input-dependent attention-score calculation $QK^T$, the corresponding operands are synchronously loaded and processed by the photonic core. The complete 16 × 16 tiling strategy is described in Supplementary text S2.4.

During inference, the programmed weight matrices remain unchanged while different input tokens and samples are continuously streamed into the photonic core. The high-speed lithium-niobate electro-optic input path enables rapid activation loading, whereas the weight path was updated when the computation advanced to the next matrix tile. This streamed-input and weight-stationary operation enables repeated reuse of the finite-size photonic core across large neural-network workloads.

The hybrid photonic implementation preserved the output distribution of the electronic reference model (Fig. 5B). For the four AG News categories, the correct-class fractions for World, Sports, Business and Sci/Tech were 0.96, 0.94, 0.82 and 0.90, respectively, compared with 0.96, 0.98, 0.86 and 0.92 for the electronic model. The resulting macro-averaged diagonal accuracy was 0.905 for the photonic implementation and 0.930 for the electronic reference. The dominant confusion between Business and Sci/Tech categories was also observed in the electronic model.

## Discussion and conclusion

This work establishes a signed intensity-domain photonic computing core that enables the physical execution of gradient-based learning within an incoherent photonic architecture. By implementing four-quadrant matrix multiplication between signed inputs and signed weights within the same photonic computing array, this architecture can not only process signed activation variables during forward propagation, but also process the signed error signals and gradient-related quantities generated during backpropagation. This capability provides a universal physical computing platform for forward computation and gradient propagation without requiring separate positive and negative photonic weight arrays. The calibrated 16 × 16 photonic core maintains stable precision across 256 parallel channels, preserves approximately 10-bit precision when extended to larger-scale matrix computation through matrix tiling, and maintains stable performance during successive computation cycles. Beyond matrix computation, the same physical core further implements the computational processes required during training, including PReLU activation function and its derivative, MSE error-signal formation, and error backpropagation. The XOR experiment places these computational processes within a complete physical training loop, whereas the BERT-mini mapping further extends this signed computational mechanism to the repeatedly occurring linear computational pathways in modern neural network models. These results establish a reusable signed linear-algebra photonic computing core capable of performing MAC operations and the signed intermediate variable computations required during backpropagation in hybrid photonic systems.

This physical architecture inherently aligns hardware computation characteristics with neural network dataflow characteristics (*37-38*). During computation, activation data need to be continuously refreshed, whereas weight states can be reused across large numbers of input vectors and reconfigured only during parameter updates. The lithium niobate electro-optic platform naturally supports this asymmetric dataflow mode: the front-end data pathway utilizes high-speed electro-optic modulation for rapid data loading, while the back-end weight pathway maintains matrix states through stable low-update-rate programming. As matrix dimensions increase, programmed weight states can be reused by larger-scale computational workloads. And combined with the advantage that the lithium niobate platform employs electro-optic tuning and avoids heating power consumption and thermal crosstalk, energy efficiency is predicted to improve with matrix size under the stated device-load model (Fig. S9).

Extending photonic processors toward larger-scale matrix computation requires overcoming the limitations of conventional lithographic processes on chip area. Future photonic accelerators for artificial intelligence computing are expected to exceed the area limitation of a single deep ultraviolet (DUV) full-reticle exposure field. The PLACE fabrication process, which enables repeated photonic structures through large-area direct writing and chemo-mechanical pattern transfer, provides a fabrication pathway for scaling interferometric structures and electro-optic modulation units beyond a single exposure field (Fig. S1). Combined with photonic–electronic co-design, this fabrication approach can support the development of larger-scale photonic computing architectures.

The wavelength domain provides a further degree of freedom for extending signed photonic computation. In the current architecture, each signed computational channel is encoded by a complementary wavelength pair, and we experimentally tested multiple wavelength combinations spanning 1546.7–1553.0 nm, all of which exhibited calibratable wavelength-dependent transmission responses (Fig. S8). Although voltage offsets exist between different wavelengths and require independent calibration for different wavelength pairs, these experiments provide a device-level basis for evaluating wavelength-division-multiplexed extensions. Combined with high-performance optical frequency comb sources (*39-43*), multiple calibrated wavelength pairs could be used to further increase the number of parallel signed computational channels while maintaining differential signed computation capability.

The characteristic and performance of ZEN-2 shown above clearly indicates a feasible pathway toward bridging universal photonic computing of high scalability and deep learning by enabling real-world data processing, physical training, and large-scale matrix computation within an incoherent signed photonic architecture.

**Acknowledgments**

The work was supported by the Quantum Science and Technology-National Science and Technology Major Project (2021ZD0301403). National Key R&D Program of China (2025YFF0524600). National Natural Science Foundation of China (12192251, 12334014, 62335019, 12134001, 12304418, 12474378).

**Funding:** Quantum Science and Technology-National Science and Technology Major Project (2021ZD0301403). National Key R&D Program of China (2025YFF0524600). National Natural Science Foundation of China (12192251, 12334014, 62335019, 12134001, 12304418, 12474378).

**Author contributions:** YR and YZ conceived the study and designed the photonic architecture and chip. YR, RL and YZ fabricated the devices. YR and YZ developed the calibration and computational mapping methods, built the experimental setup, and performed the characterization, training and inference experiments. YR and YZ developed the experimental control and data acquisition. YR and YZ analyzed the data, conducted theoretical and energy-consumption analyses, prepared the figures, and drafted the manuscript. MW and YC guided manuscript revision. YC supervised the project and secured funding. All authors discussed the results and reviewed and edited the manuscript.

**Competing interests:** YR, YZ and YC are preparing a patent application covering the photonic architecture described in this work. The authors declare no other competing interests.

**Data, code, and materials availability:** The data that support the findings of this study are available from the corresponding author upon reasonable request.

**Supplementary Materials**

Materials and Methods
Supplementary Text
Figs. S1 to S9
Tables S1

## Figure legends

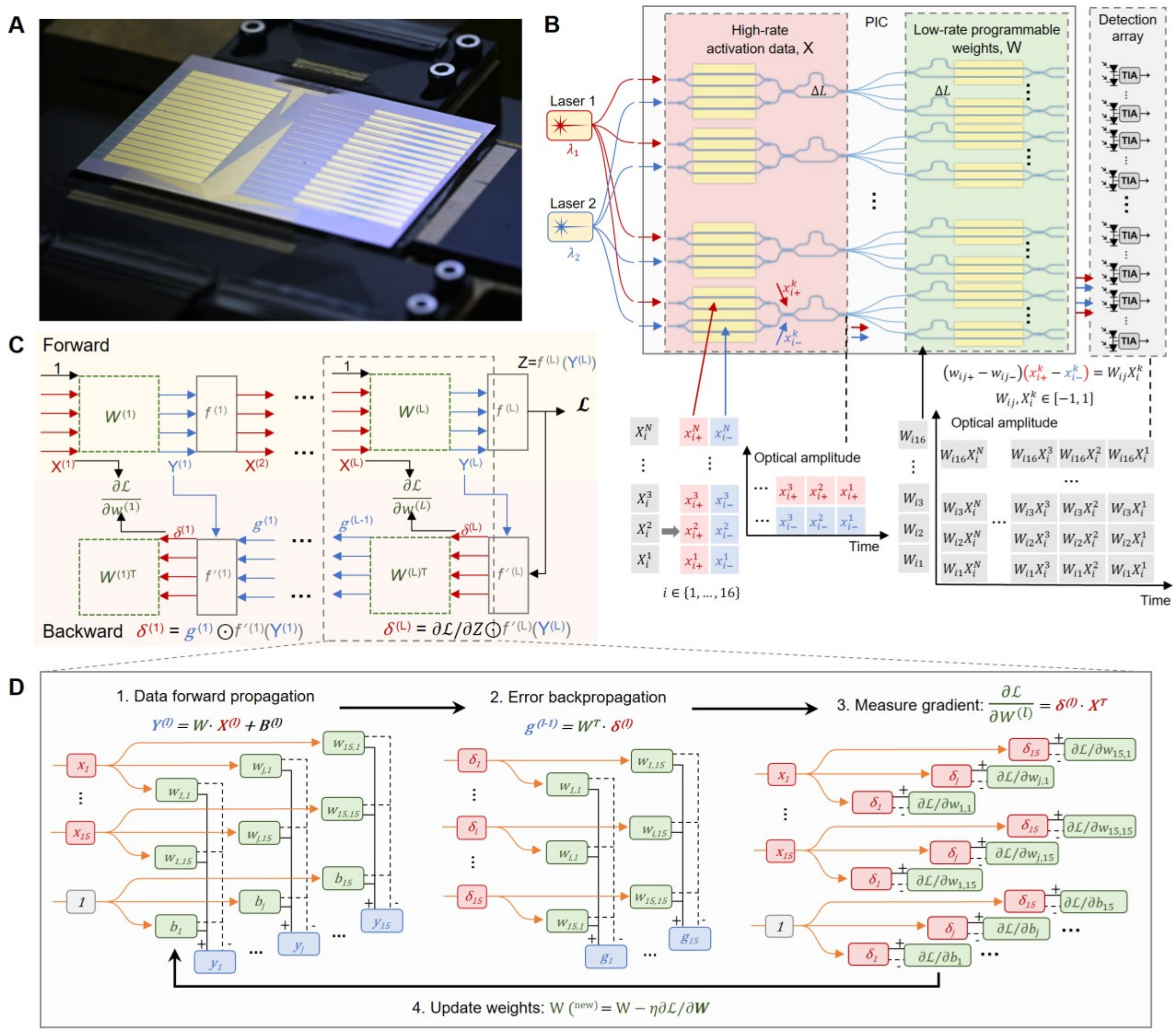

**Fig. 1. Signed incoherent photonic matrix core and in situ backpropagation dataflow.** (**A**) Photograph of the TFLN chip containing the 16 × 16 matrix core. (**B**) Dual-wavelength differential representation of signed data and separation between the high-rate input path and the lower-update-rate programmable weight path. (**C**) Layer-wise forward propagation, local transformation, error propagation and gradient relation. (**D**) Mapping of forward, backward and gradient-related operations onto the physical chip.

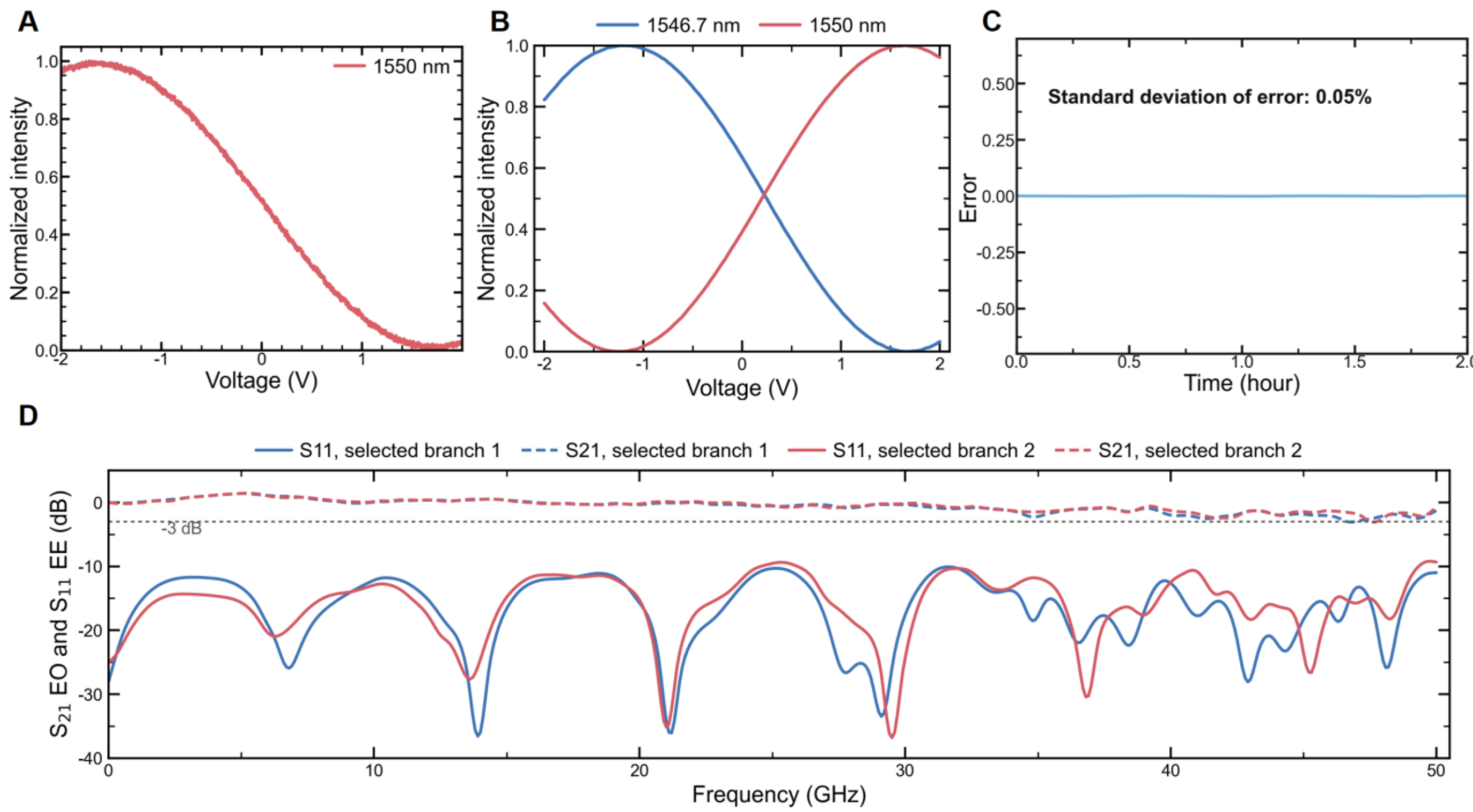

**Fig. 2. Electro-optic characterization of the high-speed input and programmable weight paths. (A)** Normalized transmission response of a representative input modulator at 1550 nm, with a fitted half-wave voltage of 3.36 V. **(B)** Normalized transmission responses of the two output ports of a representative weight modulator at 1546.7 and 1550 nm. The shared weighting MZI exhibits complementary modulation responses between the two wavelength channels, with an effective π phase spectral shift between the two transfer characteristics. **(C)** Long-term output stability measurement of the lithium-niobate electro-optic modulator under small-signal operation. The output response remains stable during the waiting interval after high-speed input modulation, showing no observable electro-optic drift during the low-update-rate weight operation. **(D)** Measured electro-optic transmission (*S*21) and electrical input reflection (*S11*).

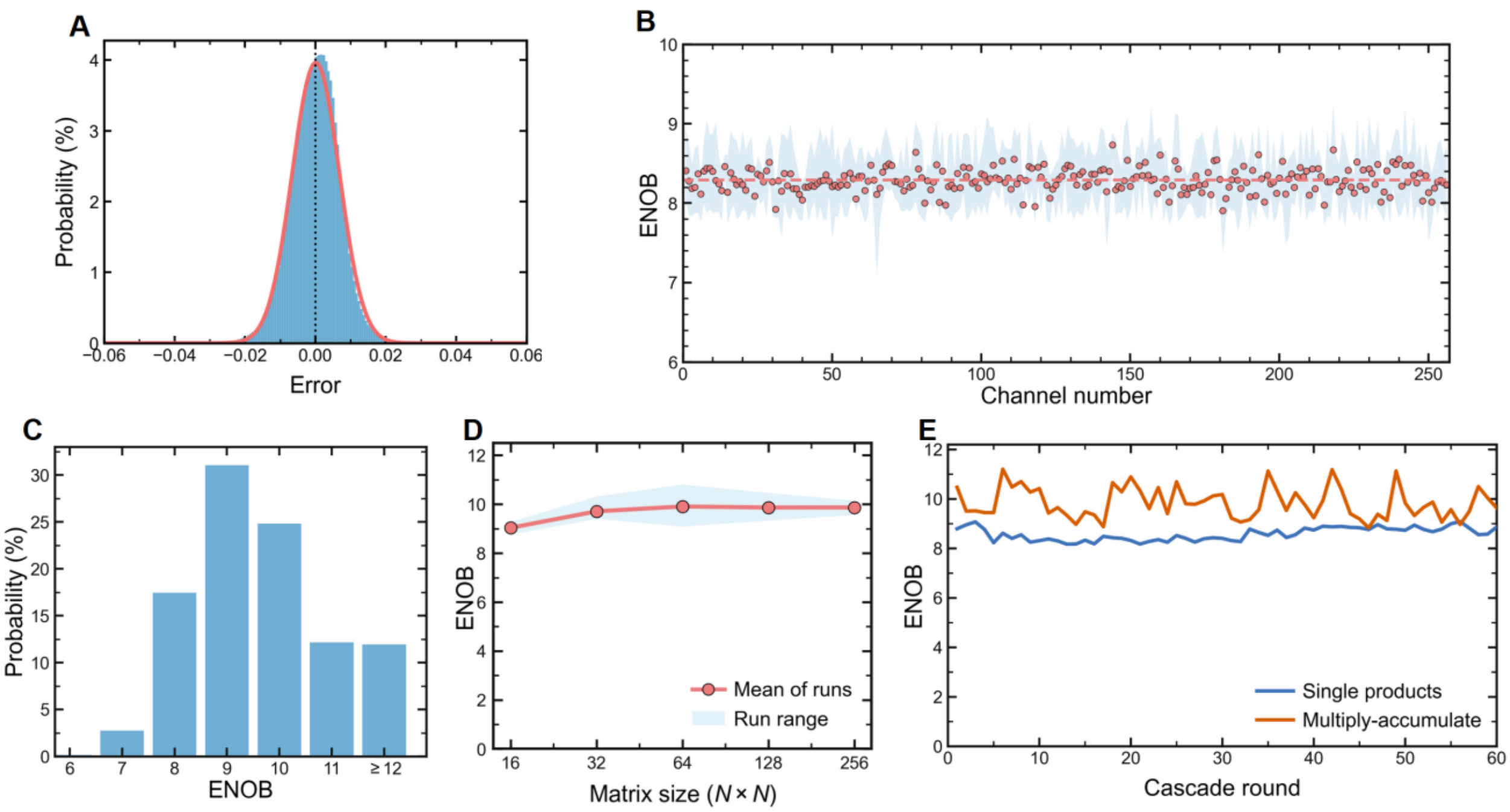

**Fig. 3. Precision of signed multiplication and matrix operations. (A)** Distribution of signed-product error measured over 1,280 independent datasets, each containing 1,000 input–weight combinations, totaling 1.28 million individual measurements. The red curve shows the fitted error distribution, and the dashed line marks zero error. **(B)** Effective number of bits (ENOB) measured across the 256 computing channels. Red points indicate the mean precision for each channel, and the shaded region shows the measured variation among repeated datasets. **(C)** Pointwise error-equivalent precision distribution for the best 1,000-point set. **(D)** ENOB of accumulated matrix outputs as the matrix dimension increases from 16 × 16 to 256 × 256 through tiled execution of the 16 × 16 core. Points show the mean of five measurements at each matrix size, and the shaded region denotes the range of the maximum and minimum ENOB. **(E)** Precision during 60 repeated 64-input computation cycles. Blue and orange traces show the ENOB of the individual signed products and the corresponding multiply–accumulate outputs, respectively.

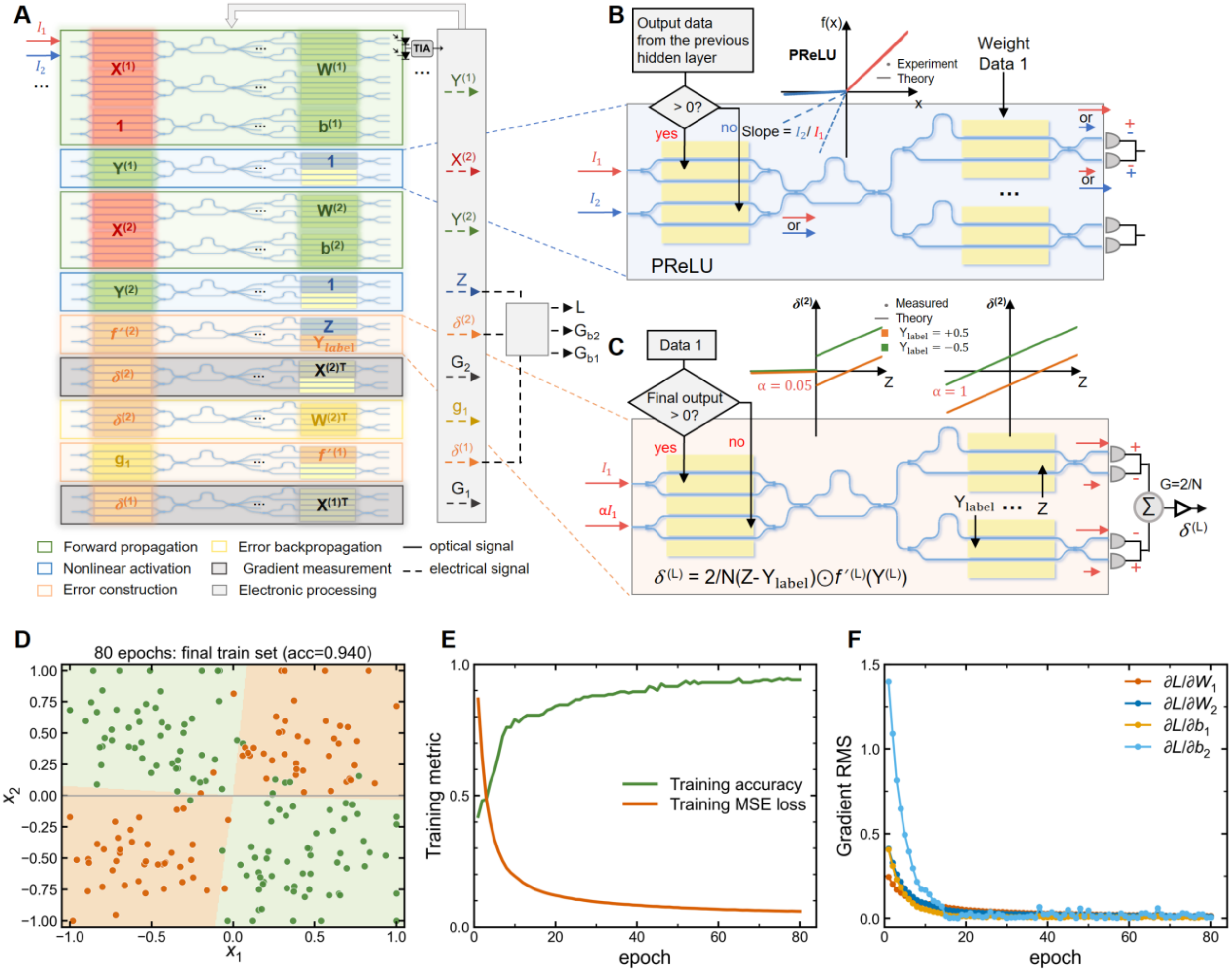

**Fig. 4. Closed-loop in situ training of a nonlinear XOR classifier with the photonic core. (A)** Schematic of the complete training procedure mapped onto the photonic processor, including forward propagation, nonlinear activation, error construction, error backpropagation, and gradient measurement. Solid and dashed lines denote optical and electrical signal paths, respectively. **(B)** Photonic implementation of the PReLU activation. According to the data sign of the previous hidden layer output, positive values activate only $\lambda_1$ with transmission 1, whereas negative values activate only $\lambda_2$ with transmission $\alpha$. With weight transmission of 1, differential detection generates the PReLU response, where $\alpha$ is optically programmable and the measured activation response agrees with the theoretical PReLU function. **(C)** Photonic construction of the output error signal for the mean-squared-error objective. The sign of the output-layer pre-activation selects the wavelength branch with a relative optical intensity of 1 or $\alpha$, thereby implementing the local activation derivative. The two weight modulators load the output value $Z$ and target label $Y_{label}$, and differential detection followed by electrical summation and scaling by $2/N$ generates the error signal required for backpropagation. **(D)** Final classification map of the noisy XOR dataset after 80 training epochs, with a training accuracy of 0.940. **(E)** Training accuracy and mean-squared-error loss as a function of epoch. **(F)** Root-mean-square magnitude of the gradient groups $\mathbf{G}_1$, $\mathbf{G}_2$, $\mathbf{G}_{b_1}$ and $\mathbf{G}_{b_2}$, defined as the gradients of $\mathcal{L}$ with respect to $\mathbf{W}^{(1)}$, $\mathbf{W}^{(2)}$, $\mathbf{b}^{(1)}$ and $\mathbf{b}^{(2)}$, respectively.

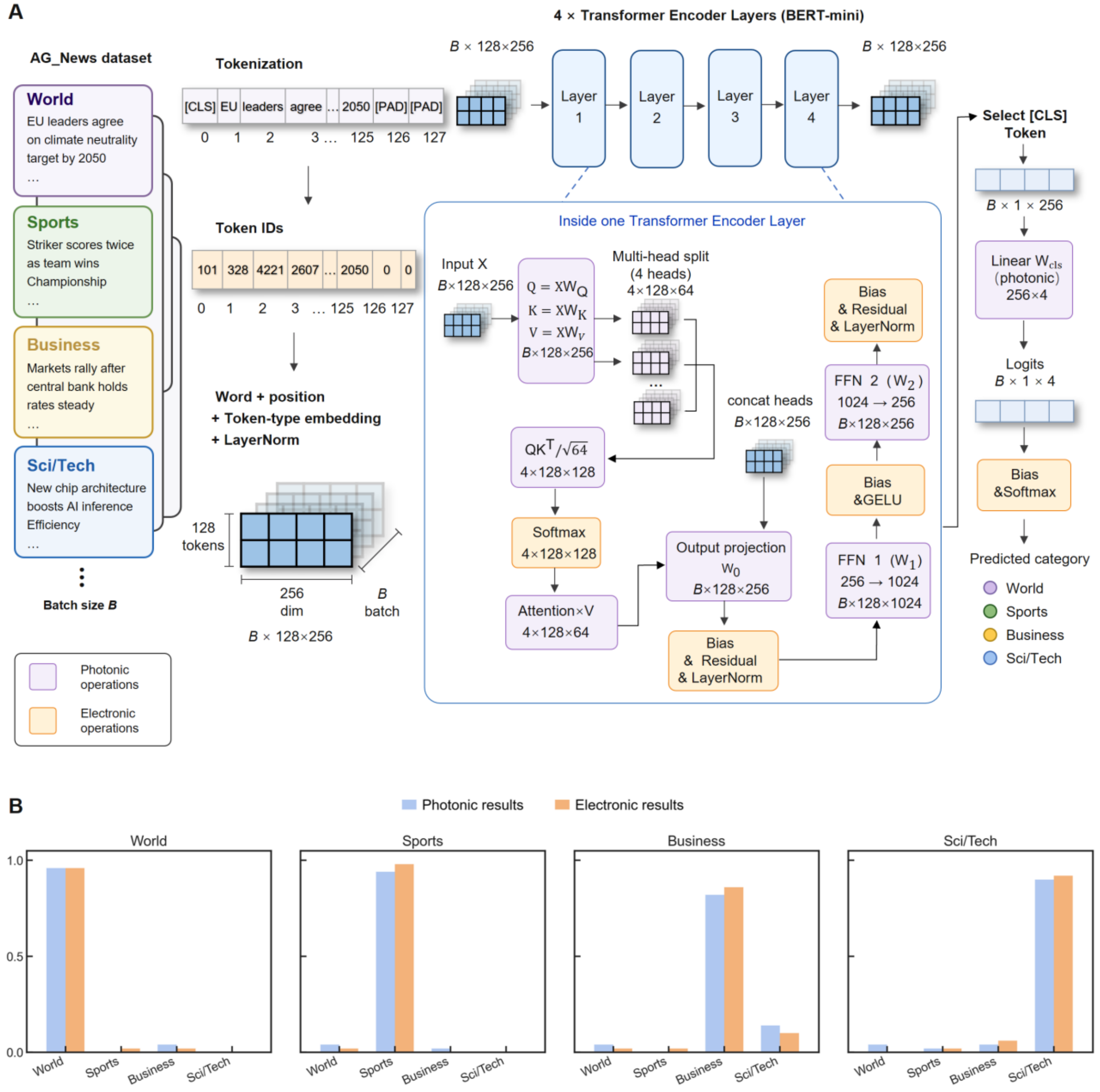

**Fig. 5. Transformer semantic classification using photonic signed linear operations. (A)** Hybrid photonic–electronic workflow for AG News classification using a four-layer BERT-mini encoder. Input text is tokenized and converted into hidden representations of size $B \times 128 \times 256$. Across the encoder and classification head, the photonic core executes the signed matrix operations associated with the query, key and value projections, attention score computation, multiplication of the attention probabilities by the value matrix, attention output projection, feed-forward layers, and classification head through tiled $16 \times 16$ matrix multiplication. Electronic processing performs tokenization, embedding lookup, bias addition, residual connection, softmax, GELU activation and layer normalization. **(B)** Comparison of class-wise output distributions between the photonic implementation and the electronic reference model for the four AG News categories. Bars represent the fractions of samples assigned to each predicted category within each true class, obtained from 200 test samples.

## Materials and methods

### S1.1 PLACE fabrication and electro-optic modulator process

The 16 × 16 processor was fabricated on a thin-film lithium niobate (TFLN) platform using photolithography-assisted chemo-mechanical etching (PLACE). The device was fabricated at the Chengya Zhixin Laboratory in China using a home-developed fabrication process. PLACE combines large-area femtosecond laser direct writing pattern definition with chemo-mechanical transfer of the structures into the lithium-niobate film, enabling the fabrication of low-loss ridge waveguides over the footprint required for the matrix processor.

The TFLN platform consists of a 500-nm-thick lithium-niobate layer bonded to a buried $SiO_2$ layer on a 500-μm-thick silicon substrate (NANOLN). The fabrication process employs femtosecond laser direct writing to define Cr hard masks, followed by chemo-mechanical polishing (CMP) to transfer the patterns into the TFLN layer. The complete fabrication flow is summarized in Fig. S1A and includes the following steps: (1) A 200-nm-thick chromium (Cr) film is deposited onto a commercial 4-inch thin-film lithium niobate (TFLN) wafer using magnetron sputtering. (2) Pre-designed photonic structures are patterned on the Cr film using a femtosecond laser direct writing system. (3) The patterns are transferred into the TFLN layer through chemo-mechanical polishing (CMP), enabling waveguides with ultra-low sidewall roughness and a propagation loss of approximately 0.027 dB/cm (*31-33*). (4) The residual Cr layer is removed using a chromium etchant. (5) A 0.8-μm-thick $SiO_2$ layer is deposited as the waveguide cladding layer. (6) A chromium–gold–titanium (Cr–Au–Ti) metal stack with thicknesses of 10 nm, 500 nm and 200 nm, is deposited by magnetron sputtering. (7) Electrode patterns are defined on the Ti layer using femtosecond laser lithography. (8) The exposed Au and Cr layers outside the protected electrode regions are removed by dry etching, using the patterned Ti layer as the etching mask. (9) The remaining Ti mask layer is removed by wet etching to finalize the electrode structures.

The device integrates two types of electro-optic electrodes, corresponding to the X-path input modulators and the W-path weight-programming cells (Fig.S1B). Although both electrode types utilize the Pockels effect in the lithium-niobate layer, they serve distinct electrical functions. The X-path electrodes are designed for high-speed electro-optic modulation of the wavelength-encoded activation signals and backpropagated error signals. These electrodes are impedance matched and operated as broadband electrical interfaces for rapid loading of time-varying neural network variables X. In contrast, the W-path electrodes control the transmission states of the unbalanced-arm MZI cells and define the programmed analog weight values. Once training is completed or during inference, the W-path voltages are maintained during repeated matrix operations and are updated only when the weight matrix is reconfigured. This separation between high-rate input modulation and low-update-rate weight programming follows the asymmetric data flow of neural-network computation (*37-38*), where activation variables are continuously streamed while weight states are reused across multiple operations. Because the lithium-niobate Pockels response is controlled by an applied electric field rather than resistive heating, the programmed W-path states do not require static heater power and avoid thermal crosstalk associated with thermo-optic tuning.

### S1.2 Signed multiplication, data encoding and training dataflow

#### S1.2.1 Differential representation and four-quadrant multiplication

Across the 256 channels, the two wavelengths experience slightly different losses. The signed optical variables are defined by differences between calibrated channel quantities, rather than by requiring the two wavelength-channel powers to have a constant sum. For a scalar input $x$, the non-negative optical intensities carried by wavelength channels $\lambda_1$ and $\lambda_2$ are $I_{\lambda_1}$ and $I_{\lambda_2}$, respectively. The encoded input is

$$x = \alpha_x \left( I_{\lambda_1} - I_{\lambda_2} \right) \tag{S1}$$

where $\alpha_x$ is the calibrated input scale. Similarly, the effective transmissions of one weight cell at the two wavelengths are $T_{\lambda_1}$ and $T_{\lambda_2}$, and the programmed signed weight is

$$w = \alpha_w \left( T_{\lambda_1} - T_{\lambda_2} \right) \tag{S2}$$

The four product terms separate into two optical readout groups. After balancing the optical transmission and electrical responsivity of the two readout channels, the same-sign and opposite-sign groups are

$$P_+ = G\left( I_{\lambda_1} T_{\lambda_1} + I_{\lambda_2} T_{\lambda_2} \right), \qquad P_- = G\left( I_{\lambda_1} T_{\lambda_2} + I_{\lambda_2} T_{\lambda_1} \right) \tag{S3}$$

where $G$ is the common calibrated optoelectronic gain after channel balancing. Differential readout gives

$$\begin{aligned} \Delta P &= P_+ - P_- \\ &= G\left( I_{\lambda_1} - I_{\lambda_2} \right)\left( T_{\lambda_1} - T_{\lambda_2} \right) \\ &= Kxw \end{aligned} \tag{S4}$$

Here $K = G/(\alpha_x \alpha_w)$. All optical intensities remain non-negative before detection. The sign of $xw$ arises from the ordering of the two calibrated detector outputs in the differential subtraction, and the recovered differential signal has the sign and magnitude of $xw$. After row-wise accumulation, the signed matrix operation is

$$y_j = \sum_i W_{ji} x_i + b_j, \qquad \mathrm{Y} = \mathrm{WX} + \mathrm{b} \tag{S5}$$

Here the photonic matrix core evaluates the linear term WX; the bias term is incorporated according to the layer-level implementation described in Section S2.3. Eq. (S4) makes the four-quadrant construction explicit. The terms $I_{\lambda_1} T_{\lambda_1}$ and $I_{\lambda_2} T_{\lambda_2}$ are assigned to $P_+$, whereas $I_{\lambda_1} T_{\lambda_2}$ and $I_{\lambda_2} T_{\lambda_1}$ are assigned to $P_-$. These labels identify their algebraic contribution after differential subtraction. They do not imply that any optical power is negative. If $x$ and $w$ have the same sign, $P_+ > P_-$, while $P_- > P_+$ when their signs differ. The calibration factor $K$ is removed when the detector voltage is mapped back to the normalized numerical product.

For Eq. (S5), $W_{ji}$ denotes the coefficient that couples input component $x_i$ to output row $j$, and $b_j$ denotes the associated bias. The differential readout therefore first recovers the contribution of each signed product and then accumulates the contributions addressed to the same output row. Eq. (S5) is consequently the matrix-level extension of the scalar primitive in equation (S4), rather than a separate numerical post-processing operation.

**S1.2.2 Dual-wavelength encoding and separation of fast and slow variables**

The physical organization of the signed photonic core is shown in Fig. S2. Before computation, the two wavelength channels are balanced by measuring a reference state and compensating for wavelength-dependent insertion loss, modulator response, and detector gain. The reference condition $x = 1$ and $w = 1$ defines the common scaling factor for the differential output. This calibration procedure does not require $I_{\lambda_1} + I_{\lambda_2}$ to remain constant during subsequent computation. The complementary relation between the two weight transmissions shown in the schematic is an implementation choice for the selected operating branch, not an additional constraint on the input encoding.

After calibration, the sign of a variable is determined by the relative intensity between the two calibrated wavelength channels. Specifically, $x > 0$, $x < 0$, and $x = 0$ correspond to $I_{\lambda_1} > I_{\lambda_2}$, $I_{\lambda_1} < I_{\lambda_2}$, and $I_{\lambda_1} = I_{\lambda_2}$, respectively, after intensity-to-value conversion according to Eq. (S1). Similarly, the relative transmission between $T_{\lambda_1}$ and $T_{\lambda_2}$ determines the sign of the weight according to Eq. (S2). Therefore, the two wavelength channels simultaneously encode the two non-negative components representing a single signed variable.

Activations and backpropagated errors are fast variables. Their two non-negative components co-propagate at $\lambda_1$ and $\lambda_2$ through the high-rate input section. The programmed weights are slower variables. For every matrix element, the weight-section voltage sets the calibrated transmission difference in Eq. (S2), and that state is held while a sequence of input or error vectors is streamed. The same four-quadrant multiplication mechanism is reused to measure each gradient-related product. For a product $\delta_j^{(l)} X_i^{(l)}$, one signed factor is encoded by the paired-wavelength input channels, whereas the other is mapped to the calibrated transmission difference of a programmable weight cell. The resulting differential readout provides the signed scalar product. The electronic controller subsequently assembles these measured products into the outer-product gradient and performs batch accumulation.

### S1.2.3 Layer-wise derivation and physical dataflow

This section gives the algebraic steps represented schematically in Fig. 1C, D. We use column-vector notation. In layer $l$, $\mathbf{X}^{(l)} \in \mathbb{R}^{n_{l-1}}$ is the input activation, $\mathbf{Y}^{(l)} \in \mathbb{R}^{n_l}$ is the pre-activation, $\mathbf{W}^{(l)} \in \mathbb{R}^{n_l \times n_{l-1}}$ is the programmed matrix and $\mathbf{b}^{(l)} \in \mathbb{R}^{n_l}$ is the bias. The index convention is important: $W_{ji}^{(l)}$ maps input component $i$ to output component $j$. The forward matrix operation and the following local nonlinearity are

$$\begin{aligned} \mathbf{Y}^{(l)} &= \mathbf{W}^{(l)}\mathbf{X}^{(l)} + \mathbf{b}^{(l)}, \\ Y_j^{(l)} &= \textstyle\sum_{i=1}^{n_{l-1}} W_{ji}^{(l)} X_i^{(l)} + b_j^{(l)}, \\ \mathbf{X}^{(l+1)} &= f^{(l)}\left(\mathbf{Y}^{(l)}\right). \end{aligned} \tag{S6}$$

Here $f^{(l)}$ is applied elementwise. At the final layer, $\mathbf{Z} = f^{(L)}(\mathbf{Y}^{(L)})$ is compared with the target $\mathbf{Y}_{\text{label}}$ through the scalar loss $\mathcal{L}(\mathbf{Z}, \mathbf{Y}_{\text{label}})$. Defining the error at the output pre-activation rather than directly at the output, gives the first chain-rule relation,

$$\boldsymbol{\delta}^{(L)} = \frac{\partial \mathcal{L}}{\partial \mathbf{Y}^{(L)}} = \frac{\partial \mathcal{L}}{\partial \mathbf{Z}} \odot f'^{(L)}\left(\mathbf{Y}^{(L)}\right), \qquad \delta_j^{(L)} = \frac{\partial \mathcal{L}}{\partial Z_j} f'^{(L)}\left(Y_j^{(L)}\right), \tag{S7}$$

where $\odot$ denotes elementwise multiplication. For an intermediate layer, the loss first propagates through the transpose of the linear map and is then multiplied by the derivative of the preceding local nonlinearity,

$$\begin{aligned} \mathbf{g}^{(l-1)} &= \left[\mathbf{W}^{(l)}\right]^{\mathsf{T}} \boldsymbol{\delta}^{(l)}, \\ g_i^{(l-1)} &= \textstyle\sum_{j=1}^{n_l} W_{ji}^{(l)} \delta_j^{(l)}, \\ \boldsymbol{\delta}^{(l-1)} &= \mathbf{g}^{(l-1)} \odot f'^{(l-1)}\left(\mathbf{Y}^{(l-1)}\right), \\ \delta_i^{(l-1)} &= g_i^{(l-1)} f'^{(l-1)}\left(Y_i^{(l-1)}\right). \end{aligned} \tag{S8}$$

The elementwise form in Eq. (S8) shows the origin of the transpose. During the forward pass, $W_{ji}^{(l)}$ carries the contribution from $X_i^{(l)}$ to $Y_j^{(l)}$. During the backward pass, the same coefficient multiplies the error associated with output component $j$ and contributes it to the error associated with input component $i$. To implement this operation experimentally, the voltage settings corresponding to $\left[\mathbf{W}^{(l)}\right]^{\mathsf{T}}$ are loaded into the same programmable array, and the signed error vector $\boldsymbol{\delta}^{(l)}$ is encoded through the paired-wavelength input channels. The resulting differential outputs evaluate $\left[\mathbf{W}^{(l)}\right]^{\mathsf{T}} \boldsymbol{\delta}^{(l)}$. The transposed matrix is therefore a reordering of the same trained coefficients rather than an independently trained parameter matrix.

The gradient of one matrix element follows directly by differentiating the component form of Eq. (S6),

$$
\begin{aligned}
\frac{\partial \mathcal{L}}{\partial W_{ji}^{(l)}} &= \sum_{k=1}^{n_l} \frac{\partial \mathcal{L}}{\partial Y_k^{(l)}} \frac{\partial Y_k^{(l)}}{\partial W_{ji}^{(l)}} \\
&= \sum_{k=1}^{n_l} \delta_k^{(l)} \ \mathbb{1}_{k=j} \, X_i^{(l)} \\
&= \delta_j^{(l)} X_i^{(l)}.
\end{aligned} \tag{S9}
$$

The indicator $\mathbb{1}_{k=j}$ equals one only for $k = j$, because $W_{ji}^{(l)}$ appears only in the pre-activation of output $j$. Thus, for one sample, the full gradient is the outer product $\partial \mathcal{L} / \partial \mathbf{W}^{(l)} = \boldsymbol{\delta}^{(l)} [\mathbf{X}^{(l)}]^{\mathsf{T}}$. For a batch of $M$ samples, the controller assembles the measured products into the gradient matrix, averages them according to Eq. (S10), and subsequently updates the corresponding programmed states,

$$
G_{ji}^{(l)} = \frac{1}{M} \sum_{m=1}^{M} \delta_{j,m}^{(l)} \, X_{i,m}^{(l)} = \frac{1}{M} \sum_{m=1}^{M} \left[ \frac{\partial \mathcal{L}}{\partial W_{ji}^{(l)}} \right]_m \tag{S10}
$$

Both factors in Eq. (S9), $\delta_j^{(l)}$ and $X_i^{(l)}$ can be positive or negative. The dual-wavelength representation is therefore used not only for forward matrix multiplication, but also for the error transport and gradient-related products required by the learning sequence. In the physical organization shown in Fig. 1D, the high-rate input region loads a signed activation or error vector. For forward and backward propagation, the programmable array is configured with the weight settings corresponding to $\mathbf{W}^{(l)}$ or $\left[\mathbf{W}^{(l)}\right]^{\mathsf{T}}$, respectively. For gradient-related multiplication, one of the two signed factors is encoded by the paired-wavelength input channels and the other is programmed into the selected weight cell. The differential output then provides the corresponding element $\delta_j^{(l)} X_i^{(l)}$, from which the controller assembles the outer-product gradient.

The state update can be expressed either in the normalized weight coordinate or in the device-voltage coordinate. Let $F_{ji}^{(l)}(u)$ be the calibrated monotonic transfer function of a selected weight-cell branch, so that $W_{ji}^{(l)} = F_{ji}^{(l)}(u_{ji}^{(l)})$. A weight-space update and its voltage-space chain rule are

$$
\begin{aligned}
W_{ji,\text{new}}^{(l)} &= \Pi_{[W_{\min}, W_{\max}]} \left( W_{ji}^{(l)} - \eta G_{ji}^{(l)} \right), \\
u_{ji,\text{new}}^{(l)} &= \left[ F_{ji}^{(l)} \right]^{-1} \left( W_{ji,\text{new}}^{(l)} \right), \\
\frac{\partial \mathcal{L}}{\partial u_{ji}^{(l)}} &= \frac{\partial \mathcal{L}}{\partial W_{ji}^{(l)}} \frac{\partial W_{ji}^{(l)}}{\partial u_{ji}^{(l)}}.
\end{aligned} \tag{S11}
$$

Here $\eta$ is the learning rate and $\Pi_{[W_{\min}, W_{\max}]}$ denotes projection onto the calibrated programmable interval. The first two lines describe the experimentally convenient lookup-table procedure: the desired normalized weight is calculated first and is then mapped to its voltage by the inverse calibration. The final line gives the corresponding chain-rule gradient when the optimization is formulated directly in the device coordinate. Because the forward, backward and gradient-related signed products and linear outputs are measured from the fabricated processor, the measured device response and its residual nonidealities directly enter the optimization loop. The activation, loss evaluation and update controller can be electronic, optical or hybrid; the photonic core supplies the signed linear relations in equations (S6), (S8) and (S9) and also implements certain nonlinear processes, with implementation details provided in Section S2.3.

### S1.3 Experimental set-up

Two wavelength carriers generated by a 1550 nm continuous-wave laser and a tunable external-cavity laser are polarization controlled and coupled into the device under test. The PXIe-8861 controller coordinates the PXIe-based voltage output (PXIe-6739) and data acquisition modules (PXIe-6358) through the shared PXI trigger and clock backplane. The voltage modules load the calibrated voltages for the input matrix, weight matrix and required DC bias conditions. The photodetector outputs corresponding to the two differential-detection ports are acquired and subtracted to recover the signed computation results in data acquisition modules. The measured results are transferred to the PXIe-8861 controller for calibration compensation and partial-sum accumulation. Dataflow settings for on-chip XOR training and AG News inference are detailed in the Supplementary Text S2.3 and S2.4.

### S1.4 NI-based automated calibration protocol

All accuracy measurements used synchronized National Instruments (NI) analogue output (AO, PXIe-6739) and analogue input (AI, PXIe-6358) channels. AO channels drove the input and weight modulator voltages and AI channels recorded the two photodetector outputs. The clocks were synchronized for each finite acquisition. In the reported measurements, the sampling rate was 1 MS/s and a typical block contained 1000000 samples. Every waveform included programmed voltage points, detector normalization anchors and a recovery segment. Each programmed voltage point was associated with the two corresponding detector readings.

Calibration was executed as an automated finite-state acquisition. For each requested scan, the controller first loads a voltage-state list for every AO channel, pads and repeats the finite waveform block, starts the synchronized AO/AI tasks, and records the two detector readings for every programmed voltage point. The resulting calibration record for each device stores the selected valley and peak voltages, the fitted half-wave response, the monotonic voltage lookup table (LUT), the detector normalization endpoints, the waveform mode and the quality-control outcome. The stored record is used directly to compile subsequent input and weight commands.

The calibration target was the signed differential observable, not a single-port transmission. For the intermediate normalized coordinates $q_x$ and $q_w$, the ideal two port response can be written as

$$I_+ = q_x q_w, \qquad I_- = q_x(1 - q_w), \qquad I_{diff} = I_+ - I_- = q_x(2q_w - 1). \tag{S12}$$

Here, $q_x$ is the normalized non-negative input coordinate for a single wavelength, and $q_w$ is the normalized weight-transmission coordinate. These calibration coordinates are distinct from the signed variables $x$, $w$ and the network quantities $\mathbf{X}^{(l)}$, $\mathbf{W}^{(l)}$ in Section S1.2. For the normalized complementary weight response $w = 2q_w - 1$. In the text below, the subscripts "in" and "wt" identify input and weight devices, respectively. The final signed two-wavelength operation is calibrated in the same differential detector basis as Eq.(S12).

Before the group-wise scans, the controller resets the NI tasks and applies the selected direct-current bias states. A 16 × 16 array is addressed as sixteen repeated 1 × 16 groups. Within group $r$, the upstream $X$ input modulator is parked at the specified reference state while only one downstream weight cell $W_{rj}$ at hardware address $(r, j)$ is scanned. The remaining AO channels are held at their group-specific bias or baseline states. This isolation makes the recorded two-port response attributable to the selected weight cell and preserves an identical acquisition sequence for all sixteen cells. Here, $(r, j)$ denotes the calibration group and cell address, rather than a change to the output–input index convention $W_{rj}$ used for the network matrix.

The array was treated as sixteen repeated 1×16 structures and calibrated group by group. For each group $r$, the procedure is as follows.

**1.** 1550-nm light was injected, and the upstream $X$ input modulator was set to a relatively high transmission reference state so that all sixteen downstream $W$ weight cells received adequate power.

**2.** The sixteen $W$ weight cells were calibrated sequentially. Each cell $j$ received a forward and reverse triangular AO voltage sweep spanning a local interval slightly larger than one half-wave branch. The two detector traces were acquired simultaneously and the two sweep directions were averaged after matching readings acquired at the same voltage. For a local interval $[V_\mathrm{l}, V_\mathrm{h}]$ and $M$ voltage levels, the triangular state list is

$$V_m^{\uparrow} = V_\mathrm{l} + \frac{m-1}{M-1}(V_\mathrm{h} - V_\mathrm{l}), \qquad V_m^{\downarrow} = V_\mathrm{h} - \frac{m-1}{M-1}(V_\mathrm{h} - V_\mathrm{l}), \quad m = 1, \dots, M. \tag{S13}$$

Let $P_{\pm,m}^{\uparrow}$ and $P_{\pm,m}^{\downarrow}$ be the recorded detector readings during the increasing and decreasing scans. With the acquisition-order indexing in Eq. (S13), $V_m^{\uparrow} = V_{M+1-m}^{\downarrow}$. The fitting value and the scan-direction residual at each voltage are

$$\tilde{P}_{\pm}(V_m) = \frac{P_{\pm,m}^{\uparrow} + P_{\pm,M+1-m}^{\downarrow}}{2}, \qquad d_{\pm}(V_m) = P_{\pm,m}^{\uparrow} - P_{\pm,M+1-m}^{\downarrow}. \tag{S14}$$

The bidirectional average reduces scan-direction-dependent differences between the two traces, whereas $d_{\pm}(V_m)$ provides a direct check for scan-direction dependence.

**3.** Detector offsets were removed and the two port traces were normalized over the selected local sweep. The complementary weight coordinate was calculated as

$$\bar{q_w} = \frac{1}{2}\left[\frac{P_+ - \min(P_+)}{\max(P_+) - \min(P_+)} + 1 - \frac{P_- - \min(P_-)}{\max(P_-) - \min(P_-)}\right]. \tag{S15}$$

The extrema in Eq. (S15) are taken over the selected sweep. The quantity $\bar{q_w}$ combines the two normalized detector traces and is subsequently fitted to determine a monotonic programming branch.

**4.** The transfer curve was fitted with the MZI response

$$f_{cal}(V) = c_0 + c_1\cos\left(\frac{\pi V}{V_\pi}\right) + c_2\sin\left(\frac{\pi V}{V_\pi}\right). \quad \text{(S16)}$$

The fitted extrema satisfy

$$\frac{\mathrm{d}f_{cal}}{\mathrm{d}V} = -\frac{\pi c_1}{V_\pi}\sin\left(\frac{\pi V}{V_\pi}\right) + \frac{\pi c_2}{V_\pi}\cos\left(\frac{\pi V}{V_\pi}\right) = 0. \quad \text{(S17)}$$

An adjacent valley-peak pair was selected only when it provided a monotonic programming branch and sufficient detector span. The resulting lookup table is denoted by

$$V_{wt,rj} = v_{wt,rj}(q_w), \qquad q_w \in [0,1]. \quad \text{(S18)}$$

The LUT $v_{wt,rj}$ maps a normalized transmission coordinate to a drive voltage. It is therefore an inverse calibration map, distinct from the voltage-to-signed-weight transfer function $F_{ji}^{(l)}$ in Eq. (S11). For an accepted valley-peak pair $(V_\mathrm{v}, V_\mathrm{p})$, the local normalized coordinate is

$$q_{rj}(V) = \frac{\bar{q_w}(V) - \bar{q_w}(V_v)}{\bar{q_w}(V_p) - \bar{q_w}(V_v)}, \qquad 0 \le q_{rj} \le 1. \quad \text{(S19)}$$

The labels $V_v$ and $V_p$ identify the response minimum and maximum, respectively. A decreasing branch has $V_p < V_v$. With ordered nodes $q_{w,l} = l/(N_{LUT} - 1)$, the stored implementation of Eq. (S18) is

$$v_{wt,rj}(q_{w,l}) = q_{rj}^{-1}(q_{w,l}), \qquad V_{wt,rj} = v_{wt,rj}(q_w) = \mathrm{interp}_{\mathrm{mon}}\left(\{q_{w,\ell}, v_{wt,rj}(q_{w,l})\}_{l=0}^{N_{LUT}-1}; q_w\right) \quad \text{(S20)}$$

The automatic branch-selection logic does not choose arbitrary global extrema. It first identifies adjacent fitted extrema and evaluates the half-wave separation $\Delta V_\pi = |V_p - V_v|$, the fitted contrast $C = |f_{cal}(V_p) - f_{cal}(V_v)|$, and the sign of $\mathrm{d}f_{cal}/\mathrm{d}V$ across the candidate interval. A candidate is accepted only when its separation is consistent with the measured $V_\pi$ within the configured margin, the fitted contrast is finite and non-negligible, and the response progresses monotonically from the selected valley to the selected peak. The accepted branch must cover the prescribed fraction of its fitted span and must not exhibit a substantial reverse excursion. During routine recalibration, the search is anchored to the previously validated half-wave branch. A candidate that shifts the branch centre or either endpoint beyond the permitted local range is rejected and the previous endpoints and LUT are restored. If the current branch is lost, a wide scan searches the allowed voltage range in overlapping $V_\pi$-scale windows and retains the valid window with the largest qualified contrast. This two-stage logic avoids silently changing to a different MZI period during a local recalibration.

**5.** After the sixteen weight cells were calibrated, the 1550-nm input curve was measured with the corresponding $W$ weight path set to the signed reference state $w = 2q_w - 1 = +1$. The 1546.7-nm input curve was measured with the $W$ weight path set to $w = -1$. These sweeps produced the wavelength-specific input lookup tables

$$V_{in,r,1550} = v_{in,r,1550}(q_{x,1550}), \qquad V_{in,r,1546.7} = v_{in,r,1546.7}(q_{x,1546.7}). \tag{S21}$$

Here, $q_{x,\lambda} \in [0,1]$ denotes the non-negative input coordinate for wavelength $\lambda$. The two channel coordinates encode the signed input through the calibrated differential representation in Eq. (S1); neither channel coordinate alone is the signed input $x$.

**6.** The procedure was repeated for every $1 \times 16$ group. The two wavelength rails were then balanced using their calibrated reference responses, with the signed condition $(x, w) = (1,1)$ defining the positive output scale. A final sparse grid of signed $(x, w)$ pairs was measured to refine, but not replace, the device-level monotonic LUTs.

Detector normalization is recalculated from four interleaved corner states after the X and W branches have been selected. The positive detector is referenced between $(q_x, q_w) = (0,0)$ and $(1,1)$, whereas the complementary detector is referenced between $(0,1)$ and $(1,0)$. In normalized form,

$$\tilde{P}_+ = \frac{P_+ - P_+^{(0,0)}}{P_+^{(1,1)} - P_+^{(0,0)}}, \qquad \tilde{P}_- = \frac{P_- - P_-^{(0,1)}}{P_-^{(1,0)} - P_-^{(0,1)}}. \tag{S22}$$

The superscripts in Eq. (S22) specify the calibration coordinates $(q_x, q_w)$, not the signed pair $(x, w)$. The corner sequence is repeated within the same finite acquisition block. Endpoint monotonicity and detector span are evaluated before the new detector calibration is accepted; an invalid result leaves the preceding valid normalization in place. These detector-normalization references are used together with the wavelength-balancing procedure described above.

The triangle scans supply the initial LUTs. Their remaining residual is evaluated using three classes of constant states: an input-coordinate scan with $q_w = 0.5$, a weight-coordinate scan with $q_x = 1$, and a sparse two-dimensional grid of $(q_x, q_w)$ states. From the normalized detector values, the effective input and weight coordinates are

$$q_{x,eff} = \tilde{P}_+ + \tilde{P}_-, \quad q_{w,eff} = \frac{1}{2}\left[1 + \frac{\tilde{P}_+ - \tilde{P}_-}{q_{x,eff}}\right]. \tag{S23}$$

The weight-coordinate estimate $q_{w,eff}$ is evaluated only when $q_{x,eff}$ exceeds the configured noise-floor criterion. The signed grid residual and its candidate-LUT score are

$$e_{\mathrm{diff}}(q_x, q_w) = (\tilde{P}_+ - \tilde{P}_-) - q_x(2q_w - 1), \qquad \mathcal{J} = \mathrm{RMSE}(e_{\mathrm{diff}}) + \lambda_{\mathrm{b}}|\bar{e}_{\mathrm{diff}}| + \lambda_{\infty}\max|e_{\mathrm{diff}}|. \tag{S24}$$

For a residual expressed in a normalized half-wave coordinate $q$, the local inverse-MZI correction is

$$h(q) = \frac{2}{\pi}\arcsin\sqrt{q}, \quad 0 \le q \le 1, \quad \Delta V_{\mathrm{model}} = sV_\pi\left[h(q_{\mathrm{target}}) - h(q_{\mathrm{meas}})\right], \quad s \in \{-1, +1\} \tag{S25}$$

where $s$ is the direction of the accepted monotonic branch. The symbol $q$ in Eq. (S25) is a generic normalized coordinate for the branch being refined, and $h(q)$ is dimensionless. The controller applies only a bounded fraction of the model correction,

$$\Delta V_{\mathrm{used}} = \mathrm{clip}(g\alpha_c \Delta V_{\mathrm{model}}, -\Delta V_{\mathrm{max}}, \Delta V_{\mathrm{max}}), \qquad v^{(k+1)}(q_{w,l}) = v^{(k)}(q_{w,l}) + \Delta V_{\mathrm{used}}(q_{w,l}). \tag{S26}$$

Here $g$ is a conservative update gain, $\alpha_c$ reduces the correction for low-confidence records, and $\Delta V_{\max}$ bounds the voltage displacement in one refinement round. Measured corrections are interpolated onto the LUT nodes and projected back to the accepted monotonic branch. Endpoint states provide consistency checks, whereas interior grid points determine the LUT-curvature correction. Each candidate LUT and its score $\mathcal{J}$ are retained; the controller commits the accepted candidate with the smallest score after detector, branch and monotonicity checks, rather than automatically selecting the final iteration.

The calibrated voltage commands for a requested signed product were therefore obtained from the corresponding non-negative calibration coordinates, $q_{x,\lambda}$ and $q_w = (w+1)/2$. The paired non-negative coordinates $q_{x,\lambda}$ are selected to represent the requested signed input $x$ according to Eq. (S1). The reported values are formed from the differential detector signal after wavelength balancing and detector normalization.

## Supplementary Text

### S2.1 Optical-path-difference consistency and spectral response of the weight cells

The W path contains 256 parallel weight cells, each implemented as an asymmetric MZI (AMZI). One arm is a straight reference path and the other contains cascaded S-bends that introduce the designed excess length. This construction retains a regular cell pitch across the array while providing the wavelength-dependent phase delay used by the dual-wavelength weight encoding. The relevant fabrication metric is therefore the optical path difference (OPD) of the complete AMZI, rather than the length of an isolated straight segment.

For an AMZI with a wavelength-dependent phase difference $\phi(\lambda)$, adjacent transmission maxima differ in phase by $2\pi$. For a fringe spacing FSR centered at $\lambda_c$, this condition gives $|\phi(\lambda_c + FSR) - \phi(\lambda_c)| = 2\pi$. For FSR $\ll \lambda_c$, first-order expansion gives $|d\phi/d\lambda|_{\lambda_c} FSR \approx 2\pi$. The corresponding effective group optical-path difference is therefore $\mathrm{OPD} \approx \frac{\lambda_c^2}{\mathrm{FSR}}$.

To test the reproducibility of the S-bend delay across the design set, devices with geometric arm-length differences $\Delta L_i$ were measured and fit with a through-origin model, $\mathrm{OPD}_i = n_{g,\mathrm{eff}} \Delta L_i$. The slope $n_{g,\mathrm{eff}}$ is a device-level effective group index. It includes the response of the straight waveguides, cascaded S-bends, MMI components, tapers and residual layout or fabrication variations in the full weight cell. It is not assigned as the intrinsic group index of a single straight-waveguide cross-section. The through-origin constraint reflects the design expectation that the differential optical path vanishes when the geometric excess length vanishes.

The measured OPD values follow the designed geometric arm-length difference with $n_{g,\mathrm{eff}} = 2.486$ and $R^2 = 0.994$ (Fig. S4A). The 147-$\mu$m group used for the array lies on the same trend, with a measured OPD of 366.945 $\mu$m and an OPD-to-$\Delta L$ ratio of 2.495. Figure S4B further shows the two output channels of the 147-$\mu$m group over 1545-1555 nm. Their complementary intensity fringes confirm the periodic AMZI response required to program a wavelength-dependent transmission difference. The independent temporal stability of the programmed intensity state is measured separately in Fig. 2C of the main text.

### S2.2 Error variables, RMSE-equivalent precision and encoding capacity

For a measurement indexed by $n$, let $y_n^{\mathrm{th}}$ be the theoretical signed output and $y_n^{\mathrm{meas}}$ be the calibrated measured output. The error and zero-reference RMSE are

$$e_n = y_n^{\mathrm{meas}} - y_n^{\mathrm{th}}, \qquad \mathrm{RMSE}_0 = \sqrt{\frac{1}{Q}\sum_{n=1}^{Q} e_n^2}. \tag{S27}$$

For the full signed output span $\Delta_{\mathrm{FS}}$, the manuscript reports the RMSE-equivalent precision

$$\mathrm{ENOB}_{\mathrm{RMSE}} = \log_2\left(\frac{\Delta_{\mathrm{FS}}}{\mathrm{RMSE}_0}\right). \tag{S28}$$

This is an error-span equivalent resolution and is not presented as a conventional sinusoidal signal-to-noise-and-distortion ENOB. The definition follows directly from the number of equal error intervals that fit inside the calibrated output span. A zero-reference error of $\mathrm{RMSE}_0 = \Delta_{\mathrm{FS}}/2^B$ corresponds to $2^B$ equal binary intervals, which gives Eq. (S28) after solving for $B$. For an output normalized to the signed interval $[-1,1]$, $\Delta_{\mathrm{FS}} = 2$. The zero reference retains both systematic offset and random variation in the reported error rather than removing the mean before computing the primary value. The pointwise visualization uses

$$B_n = \log_2\left(\frac{\Delta_{\mathrm{FS}}}{\max(|e_n|,\epsilon_{\mathrm{floor}})}\right), \qquad \epsilon_{\mathrm{floor}} = \Delta_{\mathrm{FS}} 2^{-12}. \tag{S29}$$

where $\epsilon_{\mathrm{floor}}$ implements the 12-bit display ceiling and prevents divergence at zero error. Pointwise values are used only for the distribution in Fig. 3C and are not averaged to define the reported device precision.

The terminal $\geq 12$-bit bin is chosen with reference to the voltage-addressing capacity of the electro-optic programming chain. For the 16-bit NI AO operated over $\pm 10$ V,

$$\Delta V_{\mathrm{DAQ}} = \frac{20\ \mathrm{V}}{2^{16}} = 3.0518 \times 10^{-4}\ \mathrm{V}. \tag{S30}$$

For a monotonic push-pull half-wave MZI branch, the equivalent intensity-level count based on the maximum local slope and the associated bit capacity are estimated as

$$L_{\mathrm{AO}} = 1 + \frac{2V_\pi}{\pi \Delta V_{\mathrm{DAQ}}}, \quad B_{\mathrm{cont}} = \log_2(L_{\mathrm{AO}}). \tag{S31}$$

This estimate follows from the largest local slope of a half-wave intensity transfer. Writing a normalized branch as $t(V) = \sin^2[\pi V/(2V_\pi)]$, its maximum magnitude is $|\mathrm{d}t/\mathrm{d}V|_{\max} = \pi/(2V_\pi)$. One AO least-significant voltage increment therefore corresponds, near quadrature, to a normalized intensity increment $\Delta t_{\max} = \pi \Delta V_{\mathrm{DAQ}}/(2V_\pi)$. Taking one interval per such increment gives $L_{\mathrm{AO}} \approx 1 + 1/\Delta t_{\max}$. The calculation is an addressability estimate of the programming chain; it does not substitute for the measured RMSE statistic in Eq. (S28). It also does not include analogue noise, DAC nonlinearity or calibration residuals.

Using the measured half-wave voltages in Fig. 2 gives 12.78, 12.59 and 12.56 bits for the 1550-nm input-modulator response and the weight-modulator responses at 1550 nm and 1546.7 nm, respectively. For the display in Fig. 3C, all pointwise values at or above 12 bits are grouped into one $\geq 12$-bit display bin. The 12-bit threshold is a display convention informed by these nominal addressability estimates and is not used to calculate the RMSE-equivalent precision reported in the main text.

For a matrix-vector product, the theoretical output and the row-normalized error are

$$z_j^{\mathrm{th}} = \sum_i W_{ji} x_i, \qquad S_j = \sum_i |W_{ji}|, \qquad e_{j,\mathrm{MVM}} = \frac{z_j^{\mathrm{meas}} - z_j^{\mathrm{th}}}{S_j}. \tag{S32}$$

Here, $i$ and $j$ index the input components and output rows, respectively, and $S_j$ is the sum of the absolute weights in output row $j$. The normalized error is defined for $S_j > 0$. Eq. (S28) applied to the RMSE of $e_{j,\mathrm{MVM}}$, defines the row-normalized matrix-vector precision, with a normalized full-scale span of $\Delta_{\mathrm{FS}} = 2$. Each matrix dimension was measured in five runs; Fig. 3D displays the mean and observed range. The same zero-reference convention is used for the cascade comparison, with the individual-product error calculated on the signed $[-1,1]$ scale and the multiply-accumulate error calculated after the row normalization in Eq. (S32). These quantities have different normalization scales and are reported as separate operational metrics rather than as a direct comparison of raw scalar and accumulated-product error.

The normalization in Eq. (S32) is set by the particular programmed row, not by a fixed matrix dimension. For $|x_i| \leq 1$, the triangle inequality gives $|z_j^{\mathrm{th}}| \leq \sum_i |W_{ji}| = S_j$. Thus $S_j$ is the maximum signed magnitude permitted by that row under the normalized input range and avoids attributing a small absolute error in a weakly weighted row to an artificially high precision. The same row-wise scale is used for the measured and theoretical outputs before the RMSE is evaluated. Positive and negative product errors can partially cancel during accumulation, this can yield a higher normalized MAC precision than the individual-product precision.

For Fig. 3B, each of the 256 measured multiplication branches was evaluated in five independent measurement sets, giving 1280 sets in total. For branch $c$, the plotted centre is the arithmetic mean $\overline{B}_c = 5^{-1} \sum_{s=1}^{5} B_{c,s}$ of its five set-level RMSE-equivalent ENOB values, where $B_{c,s}$ is calculated from Eq. (S28), using the set-level RMSE defined in Eq. (S27). The shaded interval is $[\min_s B_{c,s}, \max_s B_{c,s}]$. The plot therefore reports the repeatability of the precision metric for each physical multiplication branch.

## S2.3 XOR training workflow and parameter trajectories

### S2.3.1 Network, dual-wavelength variables and optical nonlinear function

The XOR experiment used a fully connected $2 \to 16 \to 1$ neural network. The two signed input variables, $x_1 \in [-1,1]$ and $x_2 \in [-1,1]$, are optically encoded by lithium niobate electro-optic intensity modulators using the two-wavelength differential representation described above, such that the relative intensities of the two wavelength channels determine both the magnitude and sign of each input. The two inputs are fully connected to 16 hidden neurons through the first-layer weight matrix $\mathbf{W}^{(1)} \in \mathbb{R}^{16\times 2}$, producing the pre-activations $y_1^{(1)}, \dots, y_{16}^{(1)}$. The hidden-layer outputs pass through elementwise PReLU activation to generate $x_1^{(2)}, \dots, x_{16}^{(2)}$. These activations are subsequently weighted by $\mathbf{W}^{(2)} \in \mathbb{R}^{1\times 16}$and accumulated to generate the output-layer pre-activation $y^{(2)}$. A final PReLU operation produces the network prediction $z$, from which the MSE loss is evaluated against the signed target value. This note uses row-wise batch notation, which is convenient for the experimental control sequence. For $N$ samples,

$$
\begin{aligned}
\mathbf{Y}^{(1)} &= \mathbf{X}^{(1)}\left[\mathbf{W}^{(1)}\right]^{\mathsf{T}} + \mathbf{1}_{N}\left[\mathbf{b}^{(1)}\right]^{\mathsf{T}},\\
\mathbf{X}^{(2)} &= f_{hidden}\left(\mathbf{Y}^{(1)}\right),\\
\mathbf{Y}^{(2)} &= \mathbf{X}^{(2)}\left[\mathbf{W}^{(2)}\right]^{\mathsf{T}} + \mathbf{1}_{N}\mathrm{b}^{(2)},\\
\mathbf{Z} &= f_{out}\left(\mathbf{Y}^{(2)}\right).
\end{aligned}
\tag{S33}
$$

Here $\mathbf{X}^{(1)} \in \mathbb{R}^{N\times 2}$. The matrices $\mathbf{Y}^{(1)}$ and $\mathbf{X}^{(2)}$ have dimensions $N \times 16$, whereas $\mathbf{Y}^{(2)}$ and $\mathbf{Z}$ have dimensions $N \times 1$. The vector $\mathbf{1}_{N}$ contains $N$ ones and broadcasts the bias across samples. Both $f_{hidden}$ and $f_{out}$ are elementwise PReLU functions. The experiment used $N = 200$ training samples, signed target values $-0.5$ and $+0.5$, and a hidden-layer PReLU negative-branch slope of 0.05, with the output-layer PReLU negative-branch slope set to 1. The reported curve is one closed training sequence on these samples; it is not a held-out test result or a repeated-seed estimate.

For each signed logical quantity $a$, the two non-negative components $a_{\lambda_1}$ and $a_{\lambda_2}$ were encoded and loaded simultaneously at the two wavelengths, with $a \propto a_{\lambda_1} - a_{\lambda_2}$. This encoding is retained for the forward activations and for the error signals used in the backward pass.

### S2.3.2 On-chip PReLU activation and terminal-error formation

The PReLU used in the XOR workflow follows the standard piecewise relation,

$$
\mathrm{PReLU}(y) = \begin{cases} y, & y \geq 0,\\ \alpha y, & y < 0, \end{cases} \qquad \mathrm{PReLU}'(y) = \begin{cases} 1, & y \geq 0,\\ \alpha, & y < 0. \end{cases}
\tag{S34}
$$

At $y = 0$, the derivative value of 1 is an implementation convention. Fig. S5A gives the physical PReLU routing used in the experiment. The weight state of each selected branch is fixed at unity. Let $y$ denote the output received from the preceding hidden layer. For $y \geq 0$, the $\lambda_1$ modulator is enabled and its calibrated input encodes the magnitude $|y|$, while the $\lambda_2$ modulator is disabled. For $y < 0$, the $\lambda_1$ modulator is disabled and the $\lambda_2$ modulator is enabled to encode the same magnitude $|y|$. Thus, the sign is encoded solely by the selected wavelength and the selected modulator always receives the magnitude of the preceding hidden-layer output. The two wavelength channels are calibrated with an optical-intensity ratio $I_{\lambda_2}/I_{\lambda_1} = \alpha$, so that equal-magnitude modulation on the selected wavelength implements the PReLU negative-branch slope $\alpha$. The factor $\alpha$ is therefore applied through the optical-intensity ratio rather than being applied a second time to the encoded input magnitude. The same sign-selection state provides the local derivative, 1 or $\alpha$, required in backpropagation. Fig. S5C compares the measured transfer functions with the theoretical PReLU curves for $\alpha = 0$, 0.01, 0.05, 0.1, 0.25, 0.3, 0.5 and 0.75, thereby providing a direct function-level check across the programmed slopes.

The output-layer error for the mean-squared-error objective is formed as

$$
\boldsymbol{\delta}^{(2)} = \frac{\partial \mathcal{L}}{\partial \mathbf{Y}^{(2)}} = \frac{2}{N}\left(\mathbf{Z} - \mathbf{Y}_{\mathrm{label}}\right) \odot f'_{\mathrm{out}}\left(\mathbf{Y}^{(2)}\right).
\tag{S35}
$$

Here, $N$ is the number of samples in the batch, and $\mathbf{Y}_{\text{label}} \in \mathbb{R}^{N\times 1}$ contains their signed target values. Because Eq. (S35) already includes the factor $1/N$, the parameter gradients are obtained by summing the sample contributions without further batch averaging.

Fig. S5B implements the terminal-error function using the stored final forward output $Z$ as the sign-control input. For a positive output PReLU slope, $Z$ and the output pre-activation have the same sign. Its front-end selector tests $Z \geq 0$, and programs the two $X$ modulators into complementary switch states: for $Z \geq 0$, the $I$-encoded branch is set to the bar state and the $\alpha I$-encoded branch to the cross state; for $Z < 0$, the bar and cross states are exchanged. The two corresponding optical paths are then evaluated by separate weighting operations encoding $\mathbf{Z}$ and $\mathbf{Y}_{\text{label}}$. The $\mathbf{Z}$ path is read by the conventional differential-detector ordering, whereas the $\mathbf{Y}_{\text{label}}$ path is read with the two differential-detector ports interchanged, so its electrical contribution has the opposite sign. The selected $I$ or $\alpha I$ branch supplies the local derivative factor to both paths. Electrically summing their calibrated detector outputs therefore yields $(\mathbf{Z} - \mathbf{Y}_{\text{label}}) f'_{\text{out}}\left(\mathbf{Y}^{(2)}\right)$; applying the factor $2/N$ gives the corresponding element of Eq. (S35), without multiplying by the local derivative again. Fig. S5 D, E characterize the terminal-error response for individual input–target pairs. The ordinate values are reported directly in the calibrated plotting coordinate recorded in the corresponding source-data tables. The measured functions for target labels $-0.5$ and $+0.5$ agree with the corresponding theory for $\alpha = 0.05$ and $\alpha = 1$.

**S2.3.3 Error propagation, gradients and parameter update**

For each sample $n$, the backward-propagated quantities are expressed as column vectors. The hidden-layer error is calculated as

$$\mathbf{g}_n^{(1)} = [\mathbf{W}^{(2)}]^{\mathsf{T}} \delta_n^{(2)}, \qquad \boldsymbol{\delta}_n^{(1)} = \mathbf{g}_n^{(1)} \odot f'_{\text{hidden}}\left(\boldsymbol{y}_n^{(1)}\right). \tag{S36}$$

Here, $\boldsymbol{y}_n^{(1)}$, $\mathbf{g}_n^{(1)}$, and $\boldsymbol{\delta}_n^{(1)}$ are the transposes of the $n$th rows of their corresponding batch matrices, whereas $\delta_n^{(2)}$ is the $n$th entry of the batch output-error vector $\boldsymbol{\delta}^{(2)}$. With samples arranged in rows as in Eq. (S33), the equivalent batch expression is $\mathbf{g}^{(1)} = \boldsymbol{\delta}^{(2)}\mathbf{W}^{(2)}$. The gradient groups reported in Fig. 4F and resolved in Fig. S6 are

$$\mathbf{G}_2 = \left[\boldsymbol{\delta}^{(2)}\right]^{\mathsf{T}} \mathbf{X}^{(2)}, \mathbf{G}_1 = \left[\boldsymbol{\delta}^{(1)}\right]^{\mathsf{T}} \mathbf{X}^{(1)}, \quad \mathrm{G}_{b_2} = \textstyle\sum_{n=1}^{N} \delta_n^{(2)}, \mathbf{G}_{b_1} = \textstyle\sum_{n=1}^{N} \boldsymbol{\delta}_n^{(1)}. \tag{S37}$$

The matrix products in Eq. (S37) sum the per-sample outer-product contributions over the batch. Because the factor $1/N$ is already included in $\boldsymbol{\delta}^{(2)}$ in Eq. (S35) and carried into $\boldsymbol{\delta}^{(1)}$ through Eq. (S36), no additional division by $N$ is applied.

Thus $\mathbf{G}_1 = \partial\mathcal{L}/\partial\mathbf{W}^{(1)}$, $\mathbf{G}_2 = \partial\mathcal{L}/\partial\mathbf{W}^{(2)}$, $\mathbf{G}_{b_1} = \partial\mathcal{L}/\partial\mathbf{b}^{(1)}$ and $\mathrm{G}_{b_2} = \partial\mathcal{L}/\partial \mathrm{b}^{(2)}$. The groups contain 32, 16, 16 and 1 elements, respectively. Each parameter was updated once per epoch, after accumulating the contributions from all $N$ training samples, using learning rate $\eta = 0.03$, with the values bounded to the calibrated programming interval,

$$\theta \leftarrow \text{clip}(\theta - \eta \nabla_\theta \mathcal{L}, -1, 1). \tag{S38}$$

For a group G containing $K$ elements, the root-mean-square quantity in Fig. 4F is

$$\mathrm{RMS}(\mathrm{G},t)=\sqrt{\frac{1}{K}\sum_{k=1}^{K}G_k\,(t)^2}. \tag{S39}$$

Here, $t$ denotes the epoch index, and $G_k(t)$ is a gradient element evaluated before the parameter update. The averaging in Eq. (S39) is over the elements within a gradient group and is used only to summarize their magnitudes.

### S2.3.4 Epoch-level execution sequence

Each epoch is executed as a closed sequence of calibrated signed linear and nonlinear operations. First, the normalized values of $\mathbf{W}^{(1)}$, $\mathbf{W}^{(2)}$, $\mathbf{b}^{(1)}$ and $\mathrm{b}^{(2)}$ are converted to device commands using the channel-specific LUTs described in Section S1.4. For each input sample in the batch, the paired wavelength components are loaded concurrently, and the core measures the first-layer signed linear output. The PReLU primitive defined in Section S2.3.2 is invoked to construct $\mathbf{X}^{(2)}$ using the calibrated wavelength selection and intensity ratio. The output-layer linear operation and output activation are then evaluated.

The terminal-error primitive defined in Section S2.3.2 combines the predictions $\boldsymbol{Z}$, the target labels $\mathbf{Y}_{\mathrm{label}}$, and the output-activation derivative to obtain $\boldsymbol{\delta}^{(2)}$, including the factor $2/N$ in Eq. (S35). The transposed output-layer operation produces the hidden-layer pre-derivative signal in Eq.(S36), and the hidden PReLU derivative yields $\boldsymbol{\delta}^{(1)}$. This operation uses the same programmable array with voltage settings corresponding to $[\mathbf{W}^{(2)}]^{\mathsf{T}}$. For each sample, the signed products required for the weight-gradient outer products are evaluated using the photonic core, with one factor encoded at the inputs and the other programmed into the weight units. The controller assembles these measured products and sums their contributions over the batch to obtain the two weight-gradient groups in Eq. (S37). It also sums the corresponding error signals to obtain the two bias-gradient groups and updates the normalized parameters using Eq. (S38), without an additional division by $N$. The updated values are clipped to the calibrated interval before the next epoch begins. Thus, the closed-loop program order is forward matrix multiplication, nonlinear activation and terminal-error formation, transposed error propagation, gradient formation and voltage update. In every optical multiplication, the two non-negative components of each signed variable are loaded concurrently on paired wavelengths.

### S2.3.5 Element-resolved gradients and parameter trajectories

Fig. S6 resolves the individual quantities that underlie the groupwise RMS traces in Fig. 4F. At each epoch, the gradients of the batch-mean MSE, calculated using Eq. (S37), are recorded before their corresponding parameter update. Fig. S6A contains the 32 elements of $\partial\mathcal{L}/\partial\mathbf{W}^{(1)}$, Fig. S6B contains the 16 elements of $\partial\mathcal{L}/\partial\mathbf{W}^{(2)}$, and Fig. S6C contains the 16 elements of $\partial\mathcal{L}/\partial\mathbf{b}^{(1)}$ together with $\partial\mathcal{L}/\partial b^{(2)}$. The colour bars identify the element index; the two curve styles in panel C distinguish the $\mathbf{b}^{(1)}$ and $b^{(2)}$ groups.

The gradient magnitudes generally decrease during training, with small fluctuations at later epochs. Figure 4F reports the RMS of each complete gradient group according to Eq. (S39), rather than an average over selected trajectories. Fig. S6 therefore makes explicit the element-level distribution that is compressed in the main-text summary.

Fig. S6D and E show the programmed trajectories of the 32 elements of $\mathbf{W}^{(1)}$ and the 16 elements of $\mathbf{W}^{(2)}$, respectively. Fig. S6F shows the 16 components of $\mathbf{b}^{(1)}$ together with the scalar $b^{(2)}$. All values are expressed in normalized programming coordinates and are constrained by Eq. (S38) to remain within the calibrated programming interval. Taken with the gradient panels, these trajectories document the element-resolved state evolution for the same closed-loop XOR sequence described in this Supplementary Text.

## S2.4 Transformer inference and control workflow for AG News

### S2.4.1 Hybrid partition and tensor operations

The photonic processor contains a native signed $16 \times 16$ matrix–vector multiplication (MVM) core. Matrices larger than the physical array are therefore evaluated by block decomposition while preserving the exact algebra of the original matrix multiplication. As illustrated in Fig. S7A, consider a general matrix multiplication $\mathbf{Y} = \mathbf{WX}$ where $\mathbf{W} \in \mathbb{R}^{M\times K}$, $\mathbf{X} \in \mathbb{R}^{K\times N}$, $\mathbf{Y} \in \mathbb{R}^{M\times N}$. Here, input vectors are arranged as columns, and $N$ denotes the number of input columns in the current matrix multiplication. For uniform $16 \times 16$partitioning, we define

$$A = \frac{M}{16},\ B = \frac{K}{16}. \tag{S40}$$

The weight matrix is divided into $A \times B$ submatrices,

$$\mathbf{W} = \begin{bmatrix} \mathbf{W}_{1,1} & \mathbf{W}_{1,2} & \cdots & \mathbf{W}_{1,B} \\ \mathbf{W}_{2,1} & \mathbf{W}_{2,2} & \cdots & \mathbf{W}_{2,B} \\ \vdots & \vdots & \ddots & \vdots \\ \mathbf{W}_{A,1} & \mathbf{W}_{A,2} & \cdots & \mathbf{W}_{A,B} \end{bmatrix},\ \mathbf{W}_{a,b} \in \mathbb{R}^{16\times 16}. \tag{S41}$$

Each column of $\boldsymbol{X}$ represents one input vector. The $n$-th column is partitioned along its $K$-dimensional inner dimension into $B$ consecutive 16-element vectors,

$$\mathbf{x}^{(n)} = \begin{bmatrix} \mathbf{x}_1^{(n)} \\ \mathbf{x}_2^{(n)} \\ \vdots \\ \mathbf{x}_B^{(n)} \end{bmatrix},\ \mathbf{x}_b^{(n)} \in \mathbb{R}^{16\times 1}. \tag{S42}$$

The corresponding output column is similarly divided into $A$ consecutive 16-element output vectors,

$$\mathbf{y}^{(n)} = \begin{bmatrix} \mathbf{y}_1^{(n)} \\ \mathbf{y}_2^{(n)} \\ \vdots \\ \mathbf{y}_A^{(n)} \end{bmatrix}, \ \mathbf{y}_a^{(n)} \in \mathbb{R}^{16\times 1}. \tag{S43}$$

The full matrix multiplication can then be expressed exactly as

$$\mathbf{y}_a^{(n)} = \sum_{b=1}^{B} \mathbf{W}_{a,b}\, \mathbf{x}_b^{(n)}, \quad a = 1, \dots, A, \quad n = 1, \dots, N. \tag{S44}$$

Thus, no approximation is introduced by the decomposition itself. A large matrix multiplication is converted into a sequence of native $16 \times 16$-by-$16 \times 1$ MVM operations followed by accumulation of the corresponding partial results.

**Native block operation**

Fig. S7B shows one such blockwise evaluation. For a fixed output block $a$ and input column $n$, the chip sequentially evaluates

$$\mathbf{p}_{a,b}^{(n)} = \mathbf{W}_{a,b}\mathbf{x}_b^{(n)}, \ b = 1, \dots, B. \tag{S45}$$

Here $\boldsymbol{p}_{\boldsymbol{a,b}}^{(\boldsymbol{n})} \in \mathbb{R}^{16\times 1}$ is the partial output generated by one physical $16 \times 16$ computation. The partial vectors corresponding to the same output block are accumulated as

$$\mathbf{y}_a^{(n)} = \sum_{b=1}^{B} \mathbf{p}_{a,b}^{(n)} = \sum_{b=1}^{B} \mathbf{W}_{a,b}\, \mathbf{x}_b^{(n)}. \tag{S46}$$

Each large-matrix result is therefore constructed from physically measured on-chip block outputs rather than from a numerical approximation of the large matrix. For an $M \times K$ matrix multiplied by $N$ input vectors, the complete operation requires $ABN = \frac{M}{16}\frac{K}{16}N$ native $16 \times 16$ MVM operations. For each $(a, n)$, the $B$ partial 16-element output vectors are accumulated to form the final $y_a^{(n)}$.

**Weight-stationary execution and asymmetric data rates**

Fig. S7C presents the complete temporal execution order. This scheduling scheme adopts a weight-stationary mode to fully exploit the significant difference in update rates between input data and weight data. For a fixed pair $(a, b)$, the weight block $\mathbf{W}_{a,b}$ is programmed once and remains stationary on the photonic core while a sequence of input vectors $\mathbf{x}_b^{(1)}, \mathbf{x}_b^{(2)}, \dots, \mathbf{x}_b^{(N)}$ is streamed through the input modulators at successive time slots $t_1, t_2, \dots, t_N$. The chip consequently produces $\mathbf{p}_{a,b}^{(1)}, \mathbf{p}_{a,b}^{(2)}, \dots, \mathbf{p}_{a,b}^{(N)}$ without reprogramming the weight block between consecutive input columns. After all $N$ input vectors have been processed, the processor loads the next weight block $\mathbf{W}_{a,b+1}$ and repeats the high-speed input sequence. The partial outputs obtained for different values of $b$ are accumulated into the same output vectors $\mathbf{y}_a^{(n)}$.

Once $b = 1, \dots, B$ has been completed, the procedure advances to the next output-row block $a + 1$. The corresponding execution order can be written as

for a = 1 … A

for b = 1 … B

load $\boldsymbol{W}_{\boldsymbol{a},\boldsymbol{b}}$ once

input $\boldsymbol{x}_{\boldsymbol{b}}^{(\mathbf{1})}, \boldsymbol{x}_{\boldsymbol{b}}^{(\mathbf{2})}, \cdots, \boldsymbol{x}_{\boldsymbol{b}}^{(\boldsymbol{n})}$

accumulate $\boldsymbol{p}_{\boldsymbol{a},\boldsymbol{b}}^{(\boldsymbol{n})}$ to $\boldsymbol{y}_{\boldsymbol{a}}^{(\boldsymbol{n})}$

This scheduling separates the system into a high-speed input data path and a low-speed weight-programming path. The front-end lithium-niobate modulators carry the rapidly varying input sequence and are therefore used as the high-bandwidth data interface. In contrast, the 256 weight values associated with a programmed block remain unchanged during the processing of an entire group of $N$ input vectors. The weight-modulation array therefore does not need to operate at the input symbol rate; it only needs to be updated when the computation moves from one weight block to the next.

This weight-stationary strategy reduces the number of weight-programming events by reusing each loaded block across input columns. For one multiplication $\boldsymbol{Y} = \boldsymbol{W}\boldsymbol{X}$, only $AB = \frac{MK}{256}$ distinct $16 \times 16$ weight blocks need to be loaded, whereas the photonic core performs $ABN = \frac{MKN}{256}$ block MVM operations. Each loaded weight block is therefore reused for $N$ consecutive input vectors before being replaced. The input-vector presentations occur at the fast data rate; because each input block must be reused for every output-row block $a$, the complete multiplication contains $ABN = \frac{MKN}{256}$ presentations of 16-element input vectors to the core under the execution order shown in Fig. S7C.

Thus, weight reuse reduces weight reloads without changing the required numbers of block MVM operations or input-vector presentations. This temporal reuse is particularly well matched to inference. High-rate input data are continuously mapped onto the photonic core, whereas the comparatively slow weight channel serves primarily to configure the linear transformation. The architecture therefore extends a physically compact $16 \times 16$ photonic core to matrices far larger than the native array size without requiring a proportional increase in the number of high-speed weight-driving channels.

For the schematic in Fig. S7, $M$, $K$, and $N$ are shown as multiples of 16 for regular partitioning. Strictly, the $16 \times 16$ spatial tiling requires $M$ and $K$ to be divisible by 16; the $N$ input columns are processed sequentially and may in general be arbitrary, with padding used when a particular system-level grouping requires a multiple of 16.

**S2.4.2 Optical and electrical control sequence**

Using the measurement setup in Fig. S3 and Section S1.3, for a tiled matrix multiplication $\mathbf{Y} = \mathbf{W}\mathbf{X}$, the PXIe-8861 controller decomposes the weight matrices into $16 \times 16$ blocks according to the physical dimension of the photonic core. The corresponding weight blocks and input segments are selected according to the weight-stationary sequence in Section S2.4.1. The weight block is converted into calibrated voltage values through the weight lookup table and loaded into the weight electrodes, while the corresponding input block is converted into electrical waveforms and applied to the input modulation channels. After the programmed states and input voltages are established, the two wavelength channels representing each signed variable are simultaneously applied to the photonic core.

The optical outputs are converted into electrical signals by photodetectors, and the signals from the two detection ports are subtracted electronically to obtain the signed multiplication results. The partial products generated by individual 16 × 16 operations are accumulated by the PXIe-8861 controller to reconstruct the corresponding output block. For each programmed weight block, all associated input columns are processed before the next weight block is loaded. The controller iterates through the inner-dimension blocks until the complete output tile is obtained and then proceeds to the remaining output blocks. Through this tiled execution procedure, a finite-size 16 × 16 photonic core performs larger matrix operations through repeated matrix decomposition and accumulation.

For Transformer inference, the same scheduling strategy is applied to the linear operations in the encoder layers and classification head, including query, key and value projections, attention output projection, feed-forward layers. Non-matrix operations, including softmax, residual addition, layer normalization and GELU activation, are performed electronically between successive photonic matrix operations. The row-normalized confusion matrix is calculated as $C_{ij} = \frac{N_{ij}}{\sum_{j'} N_{ij'}}$, where $N_{ij}$ represents the number of samples with true class $i$ and predicted class $j$.

**S2.5 Spectral compatibility of paired wavelength encoding**

The dual-wavelength protocol retains wavelength-domain parallelism because one signed channel is represented by a calibrated wavelength pair. Fig. S8 records the static weight-path response for fourteen wavelength channels measured from 1546.7 to 1553.0 nm. The channels form seven tested pairs with a 3.3-nm separation: (1546.7, 1550.0), (1547.2, 1550.5), (1547.7, 1551.0), (1548.2, 1551.5), (1548.7, 1552.0), (1549.2, 1552.5) and (1549.7, 1553.0) nm.

For this measurement, one wavelength was selected at a time, the input path was parked at its high-transmission reference state and the selected weight-path voltage was swept across the programmed window. Let $P_\lambda(V)$ denote the steady detector signal for wavelength $\lambda$. Each trace in Fig. S8 is plotted after its own endpoint normalization,

$$T_\lambda(V) = \frac{P_\lambda(V) - min[P_\lambda(V)]}{max[P_\lambda(V)] - min[P_\lambda(V)]}. \quad \text{(S47)}$$

Here, the minimum and maximum are evaluated over the scanned voltage range for each wavelength. The sweep therefore compares the normalized transfer characteristics and voltage offsets of the weight response, rather than absolute power at different laser wavelengths. For every scanned channel, $T_\lambda(V)$ traversed the programmable range from approximately 0 to 1 over the measured voltage window. The switching voltage varies with wavelength, as expected for an unbalanced interferometric weight cell. The wavelength-resolved transfer curves provide calibration data for paired-wavelength operation. Because both wavelengths traverse the same programmable weight cell, their responses are evaluated at a common applied voltage to determine the effective signed weight.

The measurements characterize the static weight-path responses of seven wavelength pairs across 1546.7–1553.0 nm. Because each signed channel is defined by a calibrated wavelength pair, multiple pairs can be allocated across this measured spectral band while retaining the paired differential encoding. The measured response therefore provides a device-level basis for evaluating calibration requirements for wavelength-division-multiplexed extension of the core.

## S2.6 Modulator-load power and energy estimate on chip

### S2.6.1 Device-level load calculation

This note estimates the electrical load power and programming energy of the implemented $16 \times 16$ core from its measured electro-optic parameters, calculated electrode capacitance and stated operating assumptions. The high-rate term is the power delivered to the $50\text{-}\Omega$ matched loads of the 32 input modulators, and the low-rate term is the capacitive charging energy associated with programming the 256 Pockels weight electrodes. Together, these terms quantify the electrical loading intrinsic to the matrix core.

Laser generation, optical distribution, RF-driver efficiency, DAC, ADC, clocking, packaging and memory traffic are system-integration contributions outside this device-level calculation. The following derivation therefore focuses on the core properties determined directly by the measured half-wave voltages, weight-electrode capacitance, and operating rate.

The parameters used here are listed in Table S1. The fitted 1550-nm input half-wave voltage is $V_{\pi,X} = 3.36$ V. The complementary weight-path half-wave voltages are $V_{\pi,W} = 2.95$ V. In our device, the weight-electrode capacitance per unit length is calculated as $2.9395542 \times 10^{-12}$ F/m, giving $C_W = 2.9395542 \times 10^{-12}$ F/m $\times\ 0.014$ m $= 41.154$ fF for a 14-mm electrode.

### S2.6.2 High-speed two-rail signed input load

The 32 high-speed modulators encode the two non-negative rails of the 16 signed input variables. For the energy estimation, each signed input variable $x_i \in [-1,1]$ is decomposed into two mutually exclusive non-negative components:

$$x_i = x_i^+ - x_i^-, \qquad x_i^+ = \max(x_i, 0), \qquad x_i^- = \max(-x_i, 0). \tag{S48}$$

Thus, each nonzero signed input occupies one of its two non-negative rails. For a representative uniform distribution $x_i \sim \mathcal{U}[-1,1]$, the mean-square values of the positive and negative components are

$$\mathbb{E}[(x_i^+)^2] = \mathbb{E}[(x_i^-)^2] = \frac{1}{2}\int_0^1 x^2\, dx = \frac{1}{6}. \tag{S49}$$

Let $u_k \in [0,1]$ denote the normalized non-negative component loaded on high-speed channel $k$, where $u_k = x_i^+$ or $x_i^-$. For this estimate, the independent bias electrode is assumed to set the modulator at quadrature. Using an ideal sinusoidal intensity transfer function over a monotonic operating interval, the normalized optical intensity and the voltage across the corresponding 50-Ω termination are related by

$$u_k = \frac{1}{2}\left[1 + \sin\left(\frac{\pi V_{X,k}}{V_{\pi,X}}\right)\right], \qquad V_{X,k} = \frac{V_{\pi,X}}{\pi}\arcsin(2u_k - 1). \tag{S50}$$

Here, $V_{X,k} \in [-V_{\pi,X}/2, V_{\pi,X}/2]$ is referenced to zero voltage on the high-speed electrode at quadrature. The total average high-speed termination-load power is therefore

$$P_X = \frac{1}{R_L}\sum_{k=1}^{32} \mathbb{E}[V_{X,k}^2]. \tag{S51}$$

Under the uniform signed-input distribution and the encoding in Eq. (S48), each rail is zero with probability $1/2$ and otherwise uniformly distributed on $(0,1)$. Its mean-square drive voltage is consequently

$$\mathbb{E}\left[V_{X,k}^2\right] = \frac{1}{2}\left(\frac{V_{\pi,X}}{2}\right)^2 + \frac{{V_{\pi,X}}^2}{2\pi^2}\int_0^1 [\arcsin(2u-1)]^2\, du = {V_{\pi,X}}^2\left(\frac{1}{4} - \frac{1}{\pi^2}\right). \tag{S52}$$

Using $V_{\pi,X} = 3.36V$ and $R_L = 50\Omega$ for all 32 channels gives

$$P_{X,1} = \frac{V_{\pi,X}^2}{R_L}\left(\frac{1}{4} - \frac{1}{\pi^2}\right) = 33.57 \text{ mW}, \qquad P_X = 32P_{X,1} = 1.074 \text{ W}. \tag{S53}$$

**S2.6.3 Low-rate Pockels weight programming and core operating metrics**

The low-speed Pockels-programmed weight electrodes are modeled as capacitive loads. The electrostatic energy stored in one electrode after a full-range programming excursion is

$$E_{W,\text{stored}} = \frac{1}{2}C_W V_{\pi,W}^2 = 179.07 \text{ fJ}. \tag{S54}$$

Under a conventional non-energy-recovering full charge-discharge cycle, the energy drawn from the source is

$$E_{W,\text{cycle}} = C_W V_{\pi,W}^2 = 358.14 \text{ fJ}. \tag{S55}$$

If every cell is updated through that conservative full-scale cycle, the matrix-level update energy is

$$E_{W,\text{matrix}} = 256 E_{W,\text{cycle}} = 91.68 \text{ pJ}. \tag{S56}$$

At a matrix-update frequency $f_W$, this capacitive contribution gives

$$P_W = E_{W,\text{matrix}} f_{\text{W}} = 91.68 \text{ pJ} \times f_{\text{W}}. \tag{S57}$$

The matrix-update frequency depends on the execution schedule. A programmed weight block remains fixed while its input vectors are processed, but reloading blocks during tiled execution also contributes to $f_{\mathrm{W}}$, in addition to calibration and training updates. For an illustrative stationary-weight estimate, we use $f_{\mathrm{W}} = 1\mathrm{Hz}$. The corresponding weight-programming power is $P_W = 91.68\ \mathrm{pJ} \times 1\ \mathrm{Hz} = 91.68\ \mathrm{pW}$. This capacitive contribution represents only the dynamic energy associated with weight reprogramming. The static power consumption of an ideal Pockels-effect capacitive electrode after programming is negligible.

**S2.6.4 Scaling of on-chip power consumption and energy efficiency for electro-optic and thermo-optic ONNs**

The implemented core combines a high-rate optical input path with a Pockels-programmed matrix state that remains static during operation or is updated at a much lower rate. At an input symbol rate of $R_s = 50$ Gbaud, each wavelength channel performs a 16×16 matrix-vector multiplication with 256 parallel multipliers. The two wavelength channels simultaneously process the positive and negative components of the signed inputs and weights, corresponding to two parallel optical MAC streams. Therefore, the dual-wavelength architecture provides a two-fold increase in the physical parallel computation throughput. Counting both the two wavelength-parallel MAC streams and the conventional two operations per MAC convention, the peak physical computational throughput is $R_{\mathrm{peak}} = 4mN^2R_s = 51.2$ TOPS, where m is the number of layers, N is the number of modes.

This 51.2 TOPS value represents the aggregate peak dual-wavelength physical processing rate used in the main text. Together with the separately evaluated high-rate input modulation energy and low-rate Pockels weight-programming contribution, it characterizes the implemented photonic computing core based on directly measured device parameters and the demonstrated operating rate.

The scaling of on-chip power consumption and energy efficiency with network size was further evaluated for electro-optically tuned lithium niobate optical neural networks (ONNs) and thermo-optically tuned ONNs. For the lithium niobate electro-optic ONN, the total on-chip power consumption is obtained from the high-speed input modulation and low-speed Pockels weight programming contributions described in Sections S2.6.2 and S2.6.3. Assuming that each of the $m$ physical layers has its own $2N$ input modulators and $N^2$ weight cells,

$$P_{\mathrm{EO}} = m(2NP_{X,1} + N^2C_WV_{\pi,W}^2f_{\mathrm{W}}) \tag{S58}$$

Because $N^2C_WV_{\pi,W}^2f_{\mathrm{W}} \ll 2NP_{X,1}$, the modeled EO load power is dominated by input modulation and scales approximately linearly with $N$ at fixed $m$.

For comparison, thermo-optic ONNs require continuous electrical heating to maintain the programmed phase states, resulting in a persistent power consumption for each tuning element. A typical thermo-optic phase shifter consumes approximately $P_{\mathrm{aveTO}}$~10-100 mW depending on the device design and operating conditions. Therefore, for an $m$-layer incoherent ONN containing $mN(N+1)$ thermo-optic phase shifters, the total power consumption can be expressed as

$$P_{\mathrm{TO}} = mN(N+1)P_{\mathrm{aveTO}} \tag{S59}$$

where $P_{\mathrm{aveTO}}$ is the power consumption of a single thermo-optic tuning element. Consequently, thermo-optic ONNs exhibit a quadratic increase in power consumption with the number of optical modes.

The energy efficiency is defined as the computational throughput per unit power:

$$\eta = \frac{R_{\mathrm{peak}}}{P_{Total}} \tag{S60}$$

Here, $P_{Total}$ is given by Eq. (S58) or Eq. (S59), with the respective power boundaries stated above. For the single-layer $16 \times 16$ EO core, the assumed 50-Gbaud symbol rate and the load estimate in Eq. (S53) give $\eta = 47.6\ \mathrm{TOPS/W}$, with the 1-Hz capacitive weight-programming contribution negligible. This metric is not a system-level wall-plug efficiency.

As shown in Fig. S9, the energy efficiency scaling of electro-optically and thermo-optically tuned ONNs is compared as a function of the number of optical modes and network layers. The analysis indicates that electro-optically driven ONNs, such as those based on lithium niobate Pockels modulation, exhibit increasing energy efficiency with increasing network scale due to their low-power weight retention and favorable power-scaling characteristics.

In our previous work, we demonstrated a coherent lithium niobate photonic neural network chip with a $6 \times 6$ optical computing core, referred to as ZEN-1. Building upon this platform, the present generation, ZEN-2, further expands the computational scale by introducing an incoherent signed optical computing architecture with a larger programmable array and reduced requirements for global phase stabilization. The scaling analysis in Fig. S9 highlights the potential of this electro-optic architecture for future large-scale photonic neural networks.

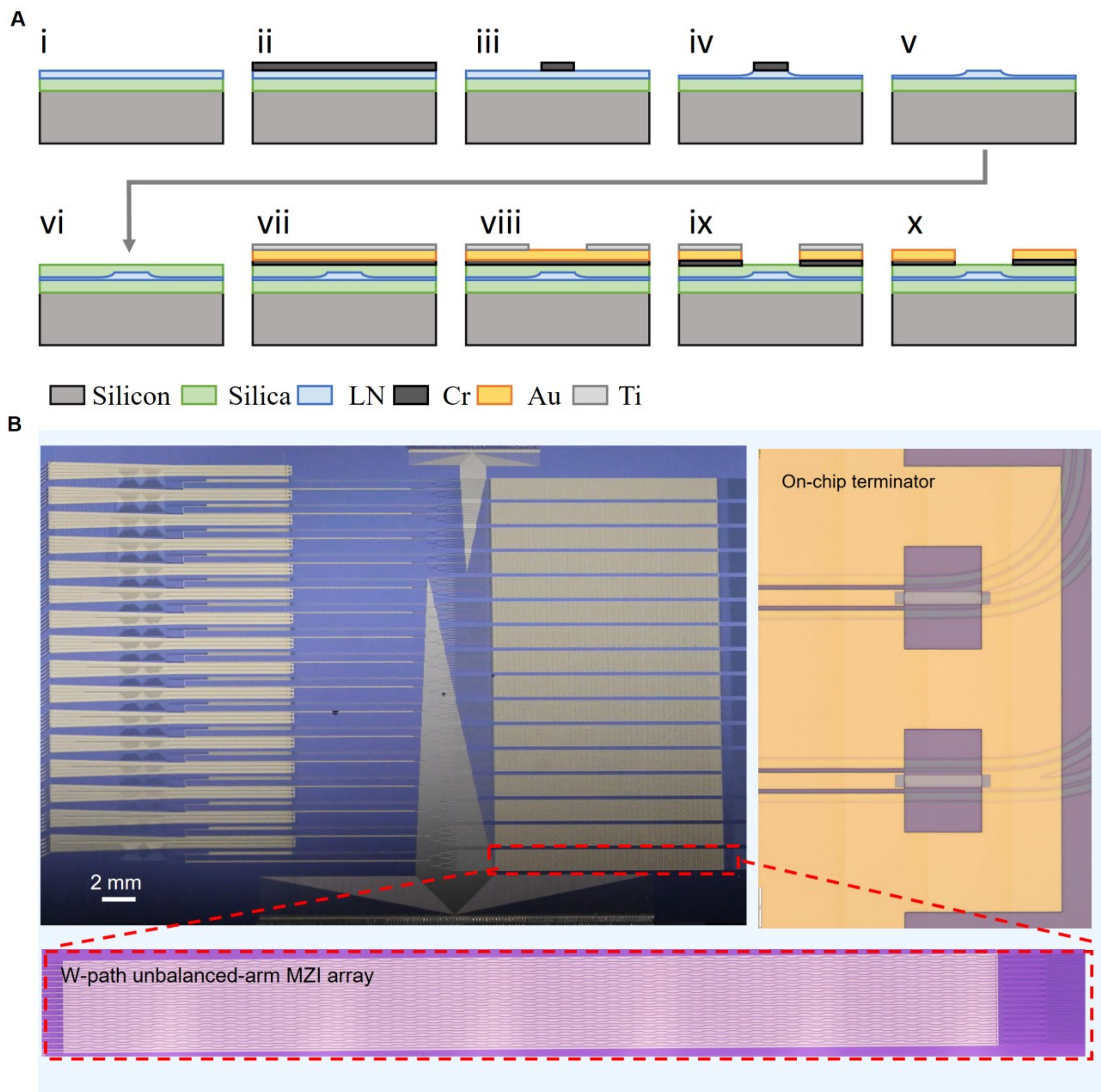

**Fig. S1 PLACE fabrication and electro-optic modulator layout.** (A) Schematic diagram of the fabrication flows of PLACE technology. (B) Optical micrograph of the fabricated 16 × 16 TFLN processor. The left panel shows the overall chip layout. The enlarged view highlights the on-chip termination loads integrated with the high-speed electro-optic input modulators. The bottom panel shows the unbalanced-arm MZI array used for the weight path.

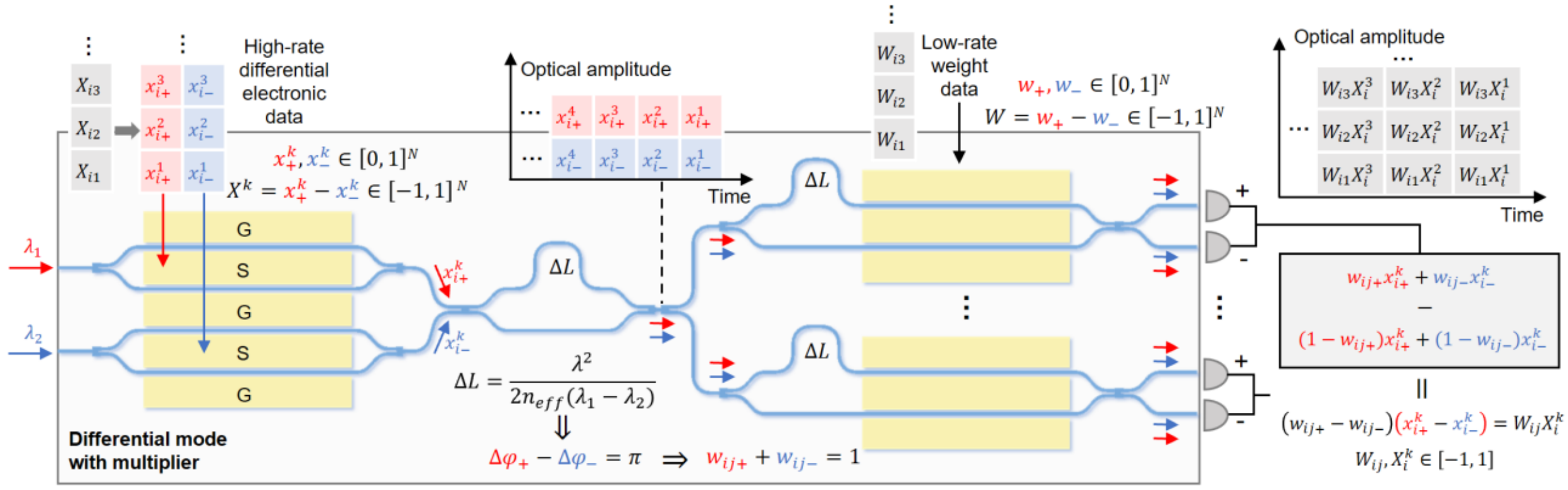

**Fig. S2 Differential dual-wavelength implementation of the signed matrix primitive.** The two non-negative components of each signed input are simultaneously encoded on two wavelength channels by the high-speed input section. An unbalanced MZI routes the pair to a shared programmable weight bank, where calibrated wavelength-dependent transmissions define the signed weight difference. Differential detection recovers the signed product and row sum. The relation $w_{ij+} + w_{ij-} = 1$ shown in the schematic is the complementary operating construction of this implementation. The general differential representation in equations (S1)–(S4) does not require the two input-channel intensities to have a constant sum.

A

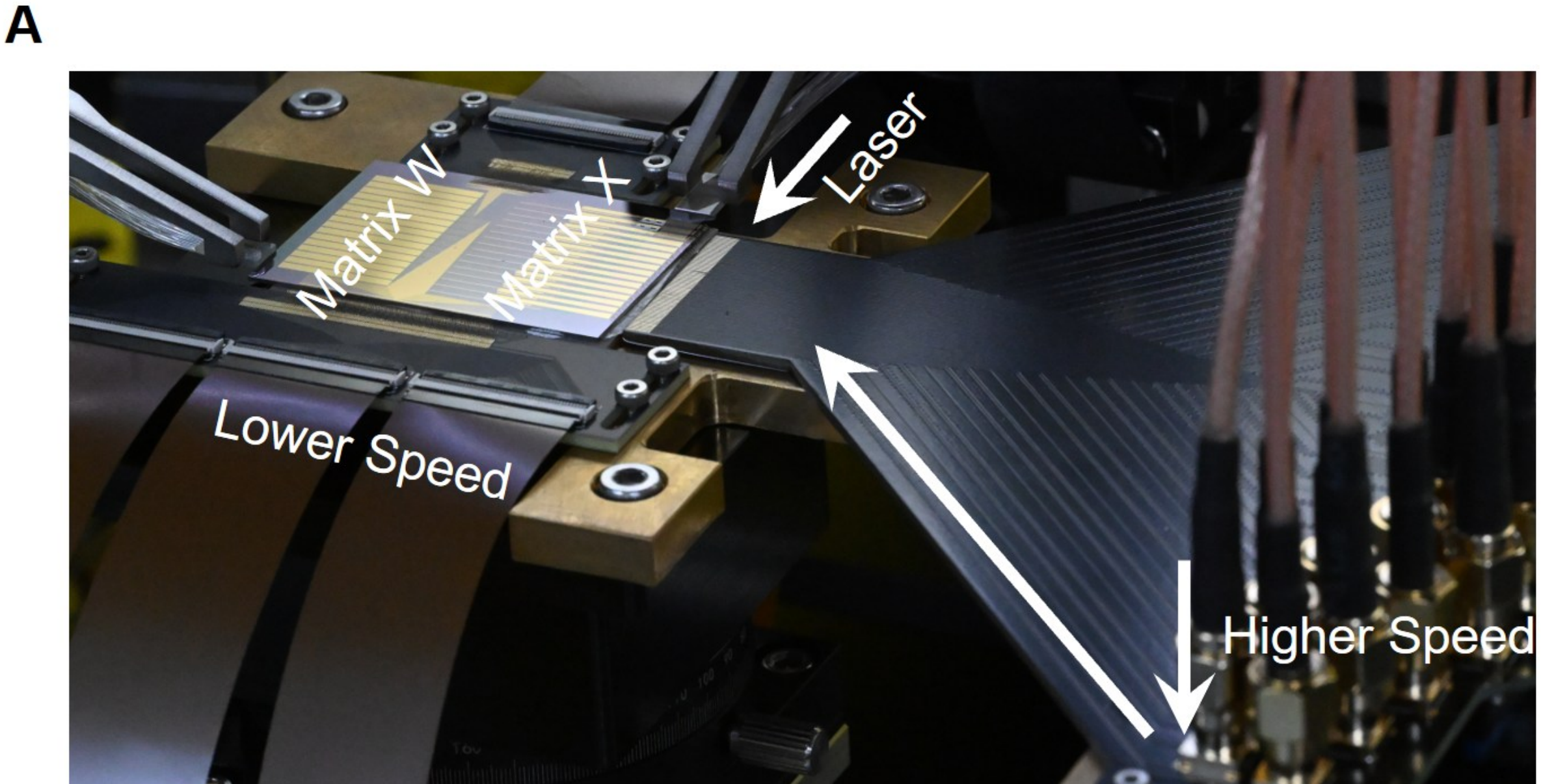

B

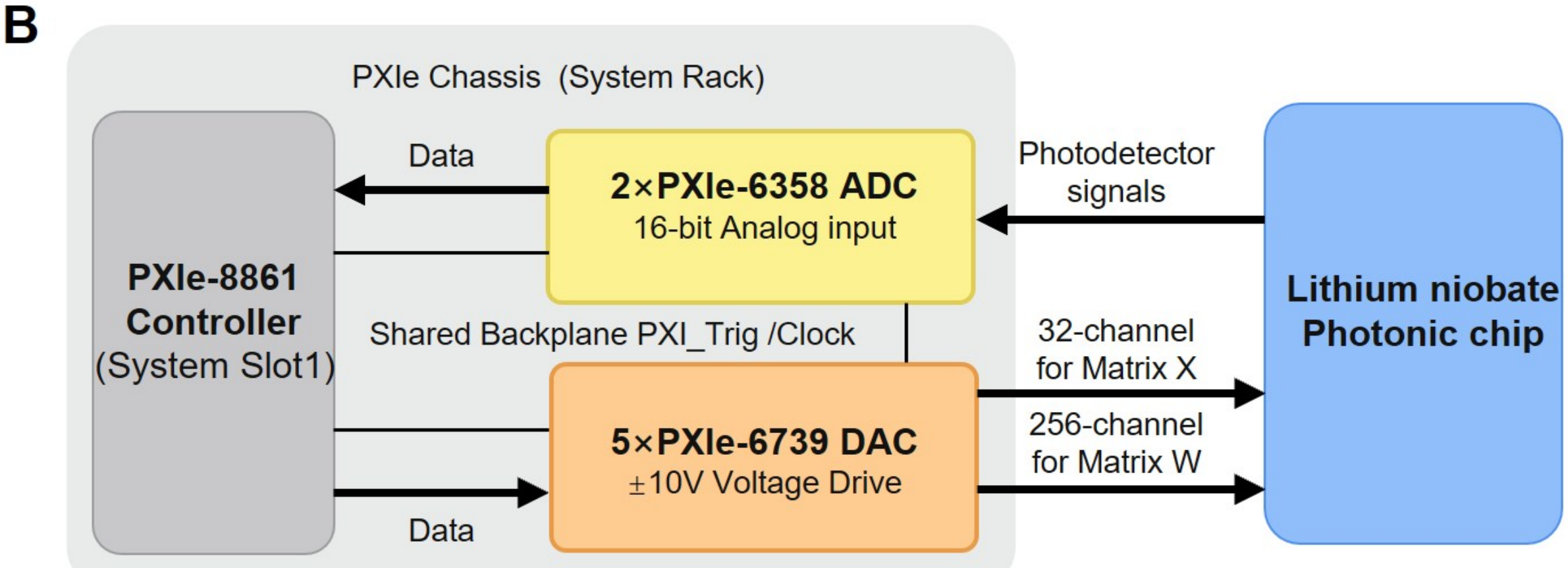

**Fig. S3. Experimental setup for the lithium-niobate photonic processor.** (A) Photograph of the packaged photonic chip and electrical interface. The chip integrates the higher speed input path (Matrix X) and programmable weight path (Matrix W). (B) Schematic of the electronic control and data acquisition architecture. A PXIe-8861 controller coordinates the PXIe-6739 DAC modules and PXIe-6358 ADC modules through the shared PXI trigger and clock backplane. The DAC modules provide voltage control for the photonic chip. The ADC modules acquire photodetector outputs after optical-to-electrical conversion and transfer the recorded signals back to the controller for subsequent processing.

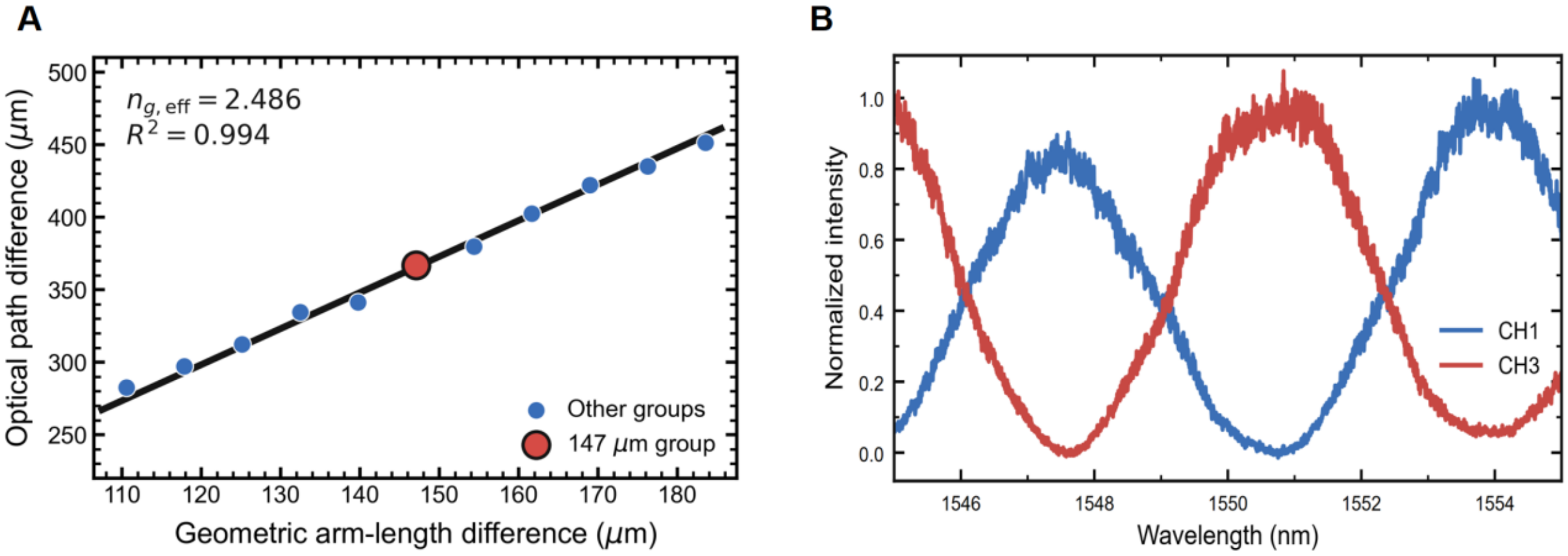

**Fig. S4 OPD calibration and spectral response of the unbalanced-arm MZI weight cell.** (A) Measured OPD as a function of the geometric arm-length difference for a set of AMZIs. The through-origin fit gives $n_{g,\mathrm{eff}} = 2.486$ and $R^2 = 0.994$; the red marker identifies the 147-$\mu$m group. (B) Wavelength-swept normalized intensities at the two output channels of the 147-$\mu$m group.

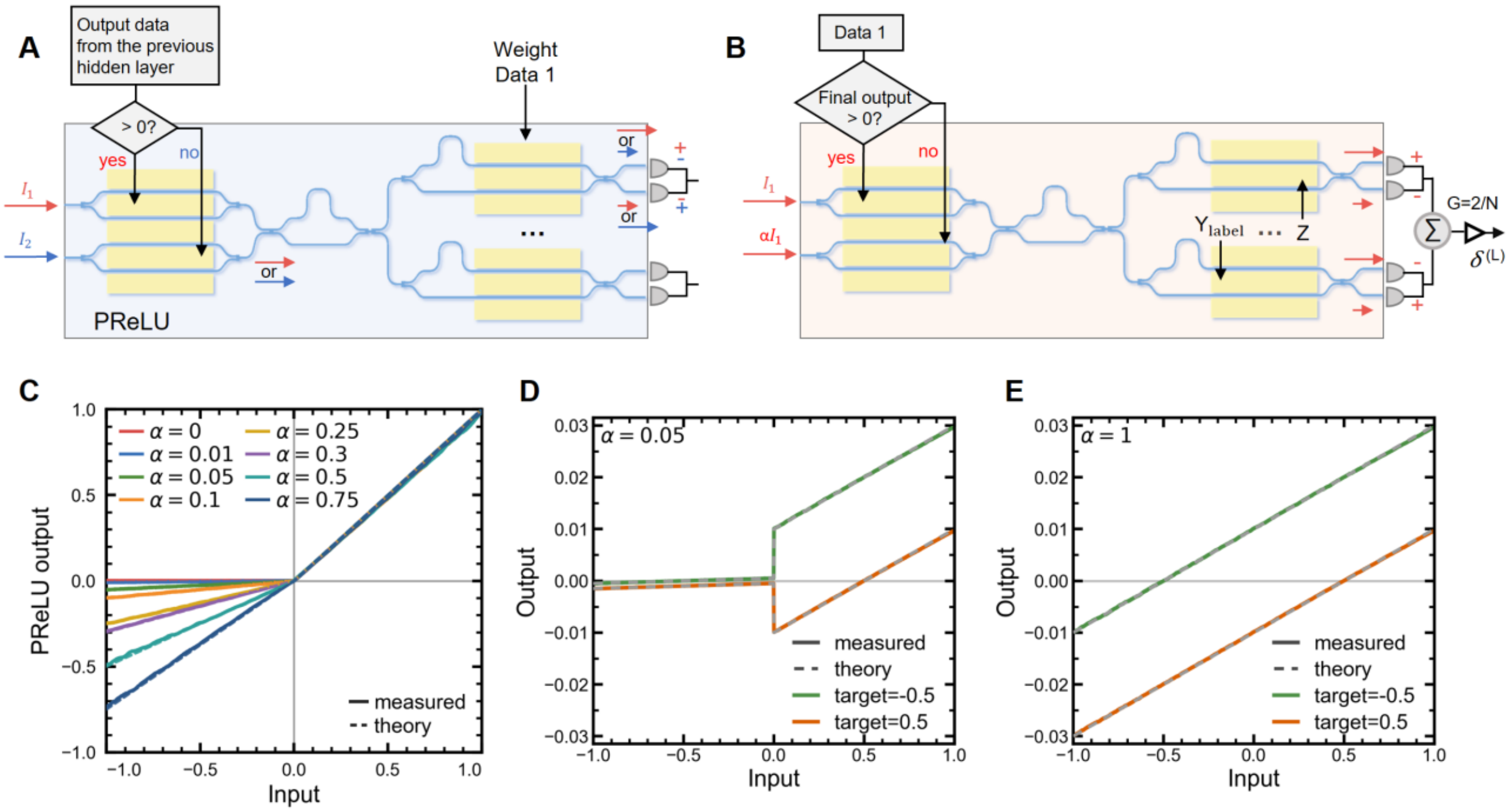

**Fig. S5 On-chip PReLU activation and terminal-error function for XOR training.** (A) Unit-weight PReLU routing. The sign of the preceding hidden-layer output selects $\lambda_1$ for $y \geq 0$ or $\lambda_2$ for $y < 0$; the enabled modulator is driven by $|y|$, and the calibrated wavelength-intensity ratio sets the negative slope $\alpha$. (B) Terminal-error construction. Stored final forward output $Z$ sets the two $X$ input modulators to complementary states: $Z \geq 0$ assigns the $I$-encoded branch to bar and the $\alpha I$-encoded branch to cross, whereas $Z < 0$ exchanges these states. The Z path uses the conventional differential-detector ordering, while the $\mathrm{Y_{label}}$ path uses the interchanged ordering; their detected outputs already include the sign-selected local derivative and are summed and multiplied by $2/N$. (C) Measured solid curves and theoretical dashed curves for eight PReLU slopes. (D-E) $N = 100$ measured and theoretical terminal-error functions for target labels $-0.5$ and $+0.5$ at $\alpha = 0.05$ and $\alpha = 1$, respectively, in the calibrated plotting coordinate.

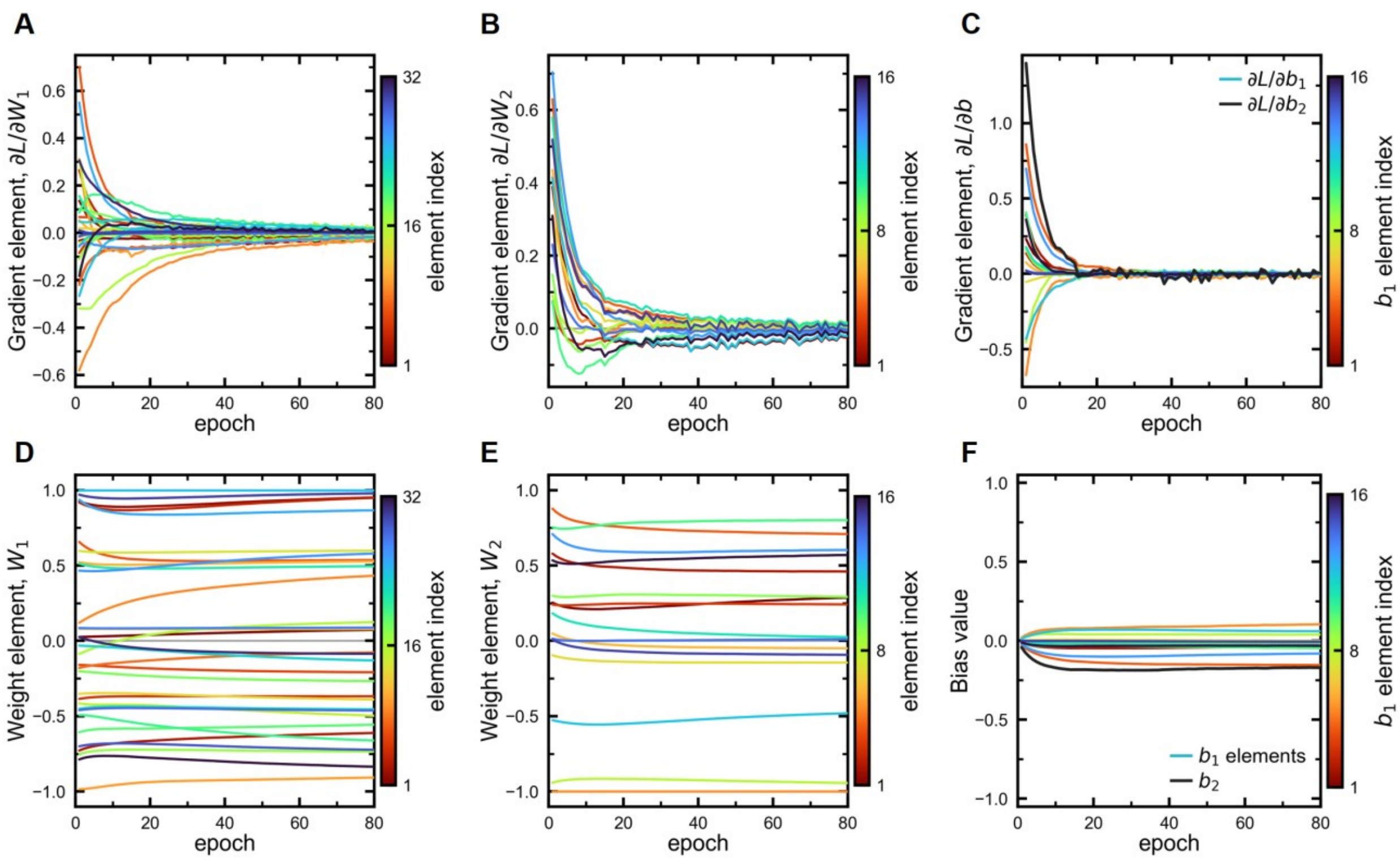

**Fig. S6 Element-resolved gradients and parameter trajectories during closed-loop XOR training.** (A-B) Trajectories of the 32 hidden-layer and 16 output-layer weight-gradient elements, respectively. (C) Trajectories of the 16 hidden-layer bias-gradient elements and the output-layer bias gradient. (D-E) Corresponding trajectories of the hidden-layer and output-layer weights. (F) Trajectories of the hidden-layer and output-layer bias values. Colour bars indicate element index; in C and F, the line styles distinguish the $\mathrm{b}^{(1)}$ and $b^{(2)}$ groups. The horizontal axis is epoch in all panels.

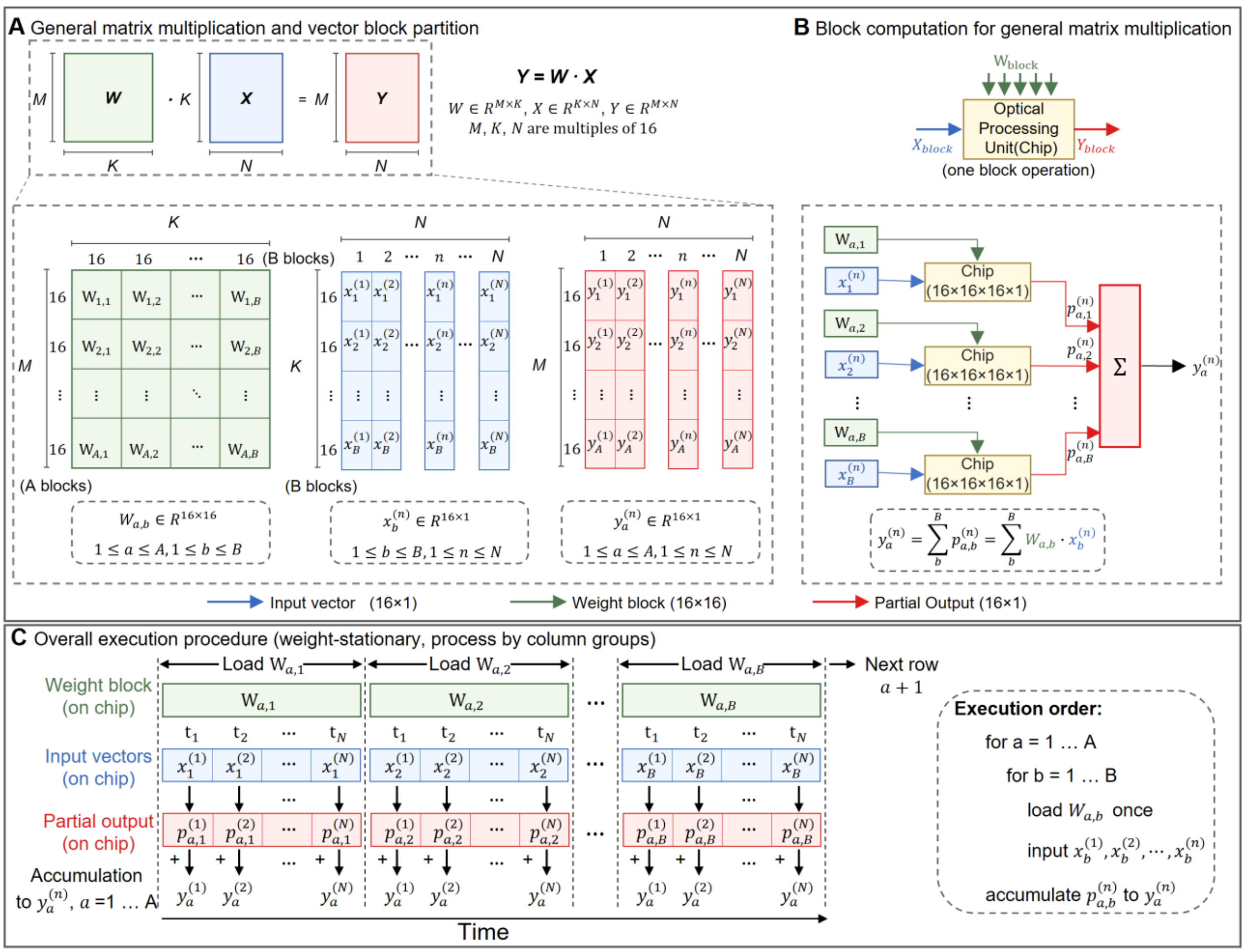

**Fig. S7. Blockwise execution of matrix multiplication beyond the native $16 \times 16$ photonic core.** (A) Decomposition of a large matrix multiplication into $16 \times 16$ matrix–vector multiplication blocks. A weight matrix $W$ and input matrix $X$ are partitioned into sub-blocks, where the output block is obtained by accumulating partial products from individual photonic core operations. (B) Execution of a single $16 \times 16$ block operation, where a programmed weight block generates a partial output with a 16-element input vector. (C) Weight-stationary dataflow for scalable computation. Weight blocks are updated at a low rate, while input vectors are continuously streamed through the high-speed input modulators. Temporal reuse of programmed weights enables matrix multiplication beyond the physical array size.

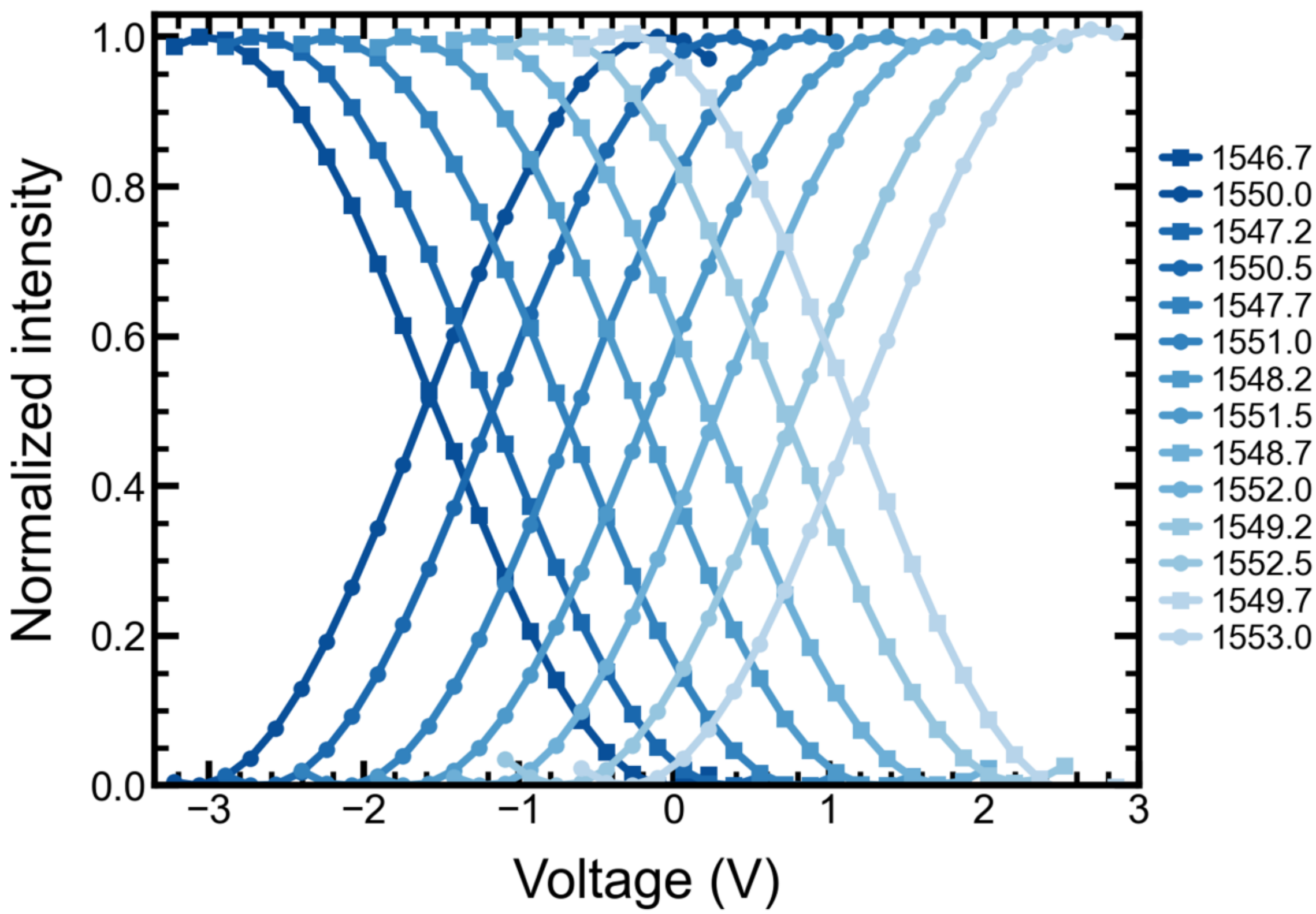

**Fig. S8. Static weight-path response across seven tested wavelength pairs.** Normalized intensity as a function of programmed voltage for fourteen wavelength channels spanning 1546.7–1553.0 nm. The traces are arranged as seven 3.3-nm wavelength pairs for differential signed encoding. Each wavelength was measured separately, and each trace was normalized independently using Eq. (S47). The measured voltage offsets characterize the wavelength dependence of the weight response and inform calibration for paired-wavelength encoding.

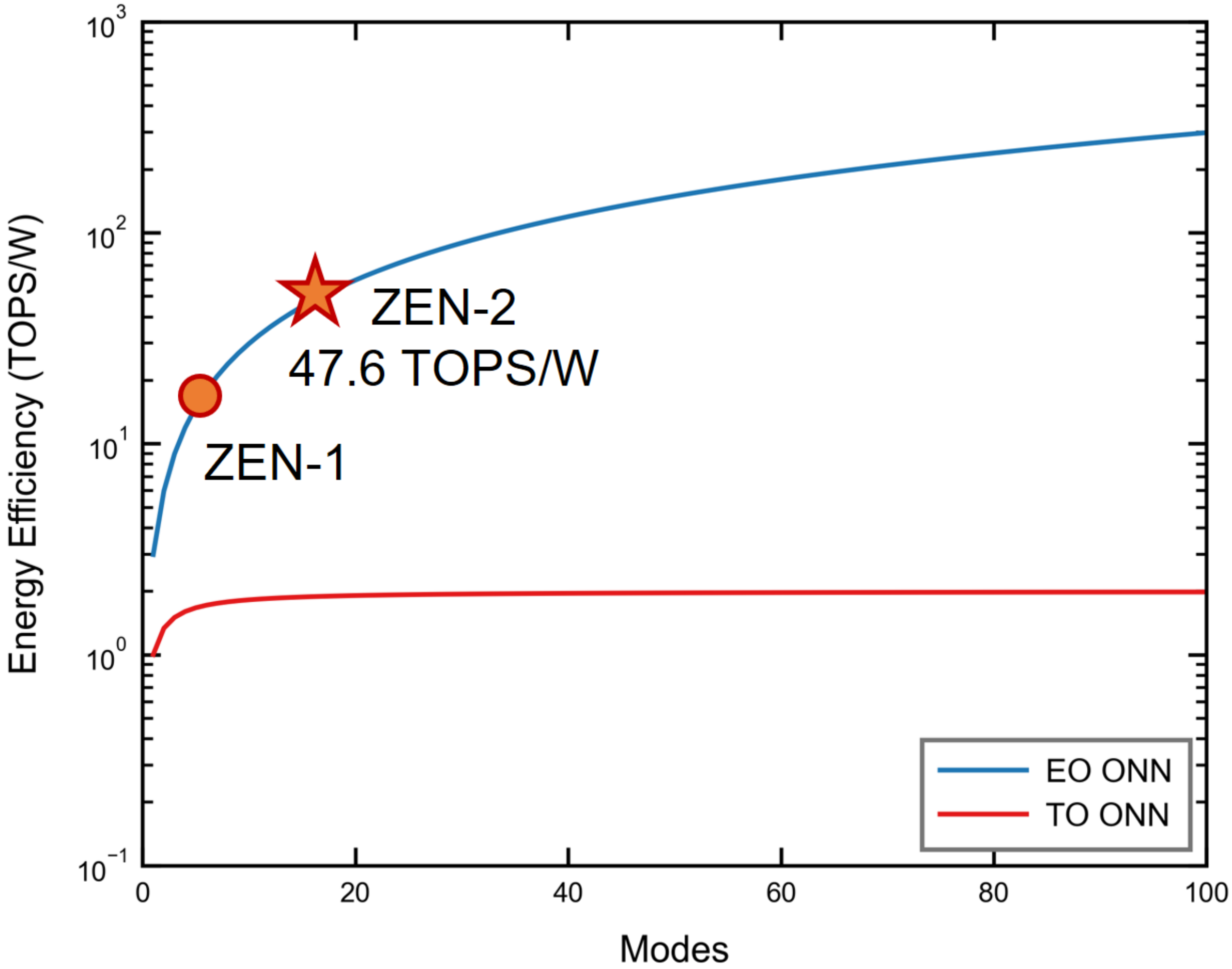

**Fig. S9. Scaling of energy efficiency for electro-optic and thermo-optic ONNs.** Energy efficiency (TOPS/W) as a function of the number of optical modes for electro-optically tuned (EO) and thermo-optically tuned (TO) ONNs under the stated device-load assumptions. The EO ONN benefits from the low-power Pockels-programmed weight states and high-speed electro-optic input modulation, resulting in a continuous improvement of energy efficiency with increasing mode number. In contrast, TO ONNs suffer from the quadratic power scaling associated with continuously powered thermal phase shifters, leading to limited energy-efficiency improvement as the network size increases. Under the assumed replication of identical layers, throughput and load power both scale with $m$, so their ratio is independent of $m$.

| Parameter | Value |
|---|---|
| Matrix size | $16 \times 16$ |
| High-speed input modulators | 32 |
| Input half-wave voltage | $V_{\pi,X} = 3.36$ V |
| Input load | $R_L = 50\ \Omega$ |
| Symbol rate | $R_s = 50$ Gbaud |
| Weight cells | 256 |
| Weight-electrode capacitance | $C_W = 41.154$ fF |
| Conservative weight half-wave voltage | $V_{\pi,W} = 2.95$ V |

**Table S1 Parameters used for the core load and energy estimate.**